%% file: main.tex
\documentclass[twocolumn,twocolappendix]{aastex63}
\usepackage{amsmath}
\usepackage{multirow}
\hypersetup{linkcolor=blue,citecolor=blue,filecolor=cyan,urlcolor=magenta}
\graphicspath{{./}}
\def \nh {$N_{\mbox{\scriptsize H}}$}
\def \cmmt {$\mbox{cm}^{-2}$}

\def \lognhcmmt {$\log\,N_{\mbox{\scriptsize H}}/\mbox{cm}^{-2}$}

\def \rpex {$R_{\mbox{\scriptsize pex}}$}

\def \ecut {$E_{\mbox{\scriptsize cut}}$}
\def \pnull {$p_{\mbox{\scriptsize null}}$}
\def \fsca {$f_{\mbox{\scriptsize s}}$}
\def \feka {{Fe\,K$\alpha$}}
\def \feew {{EW$_{\mbox{\scriptsize Fe\,K}\alpha}$}}

\def \swift {{\em Swift}}
\def \bepposax {{\em BeppoSAX}}
\def \integral {{\em INTEGRAL}}
\def \suzaku {{\em Suzaku}}
\def \nustar {{\em NuSTAR}}
\def \swiftxrt {{\em Swift}/XRT}
\def \swiftbat {{\em Swift}/BAT}

\def \rxte {{\em RXTE}}

\def \xmmnewton {{\em XMM-Newton}}
\def \pexrav {\texttt{pexrav}}

\def \xspec {\texttt{Xspec}}

\shorttitle{Median High-energy Cutoff in Seyfert II Hard X-ray Spectra}
\shortauthors{Balokovi\'{c} et al.}

\begin{document}

\phantom{.}\vspace{-0.75cm}

\title{NuSTAR Survey of Obscured Swift/BAT-selected Active Galactic Nuclei: \\ II. Median High-energy Cutoff in Seyfert II Hard X-ray Spectra}

\correspondingauthor{M. Balokovi\'{c},~\texttt{mislav.balokovic@yale.edu}}
%\email{mislav.balokovic@yale.edu}

\author[0000-0003-0476-6647]{M. Balokovi\'{c}}
\affiliation{Yale Center for Astronomy \& Astrophysics, 52 Hillhouse Avenue, New Haven, CT 06511, USA}
\affiliation{Department of Physics, Yale University, P.O. Box 2018120, New Haven, CT 06520, USA}
\affiliation{Center for Astrophysics $\vert$ Harvard \& Smithsonian, 60 Garden Street, Cambridge, MA 02138, USA}
\affiliation{Black Hole Initiative at Harvard University, 20 Garden Street, Cambridge, MA 02138, USA}

\author[0000-0003-2992-8024]{F.\,A. Harrison}
\affiliation{Cahill Center for Astronomy and Astrophysics, California Institute of Technology, 1200 East California Boulevard, Pasadena, CA 91125, USA}

\author{G. Madejski}
\affiliation{Kavli Institute for Particle Astrophysics and Cosmology, SLAC National Accelerator Laboratory, Stanford University, Stanford, CA 94305, USA}

\author[0000-0003-3451-9970]{A. Comastri}
\affiliation{INAF--OAS Osservatorio di Astrofisica e Scienza dello Spazio di Bologna, Via Gobetti 93/3, I-40129 Bologna, Italy}

\author[0000-0001-5231-2645]{C. Ricci}
\affiliation{N\'{u}cleo de Astronom\'{\i}a, Facultad de Ingenier\'{\i}a y Ciencias, Universidad Diego Portales, Av. Ej\'{e}rcito Libertador 441, Santiago, Chile}
\affiliation{Kavli Institute for Astronomy and Astrophysics, Peking University, Beijing 100871, People’s Republic of China}

\author[0000-0003-0387-1429]{A. Annuar}
\affiliation{Department of Applied Physics, Faculty of Science and Technology, Universiti Kebangsaan Malaysia, 43600 Bangi, Selangor, Malaysia}

\author[0000-0001-8128-6976]{D.\,R. Ballantyne}
\affiliation{Georgia Institute of Technology, 837 State Street, Atlanta, GA 30332, USA}

\author[0000-0001-9379-4716]{P. Boorman}
\affiliation{Department of Physics and Astronomy, University of Southampton, Southampton SO17 1BJ, UK}
\affiliation{Astronomical Institute, Academy of Sciences, Bo\v{c}n\'{\i} II 1401, CZ-14131 Prague, Czech Republic}

\author[0000-0002-0167-2453]{W.\,N. Brandt}
\affiliation{Department of Astronomy and Astrophysics, The Pennsylvania State University, University Park, PA 16802, USA}
\affiliation{Institute for Gravitation and the Cosmos, The Pennsylvania State University, University Park, PA 16802, USA}
\affiliation{Department of Physics, 104 Davey Laboratory, The Pennsylvania State University, University Park, PA 16802, USA}

\author[0000-0002-8147-2602]{M. Brightman}
\affiliation{Cahill Center for Astronomy and Astrophysics, California Institute of Technology, 1200 East California Boulevard, Pasadena, CA 91125, USA}

\author[0000-0003-3105-2615]{P. Gandhi}
\affiliation{Department of Physics and Astronomy, University of Southampton, Southampton SO17 1BJ, UK}

\author[0000-0002-3233-2451]{N. Kamraj}
\affiliation{Cahill Center for Astronomy and Astrophysics, California Institute of Technology, 1200 East California Boulevard, Pasadena, CA 91125, USA}

\author[0000-0002-7998-9581]{M.\,J. Koss}
\affiliation{Eureka Scientific Inc., 2452 Delmer St., Suite 100, Oakland, CA 94602, USA}

\author[0000-0001-5544-0749]{S. Marchesi}
\affiliation{INAF---OAS Osservatorio di Astrofisica e Scienza dello Spazio di Bologna, Via Gobetti 93/3, I-40129 Bologna, Italy}
\affiliation{Department of Physics and Astronomy, Clemson University,  Kinard Lab of Physics, Clemson, SC 29634, USA}

\author[0000-0002-2055-4946]{A. Marinucci}
\affiliation{Agenzia Spaziale Italiana (ASI) --- Unit\`{a} di Ricerca Scientifica, Via del Politecnico snc, I-00133 Roma, Italy}

\author[0000-0002-7100-9366]{A. Masini}
\affiliation{Scuola Internazionale Superiore di Studi Avanzati, Via Bonomea 265, I-34136 Trieste, Italy}

\author[0000-0002-2152-0916]{G. Matt}
\affiliation{Dipartimento di Matematica e Fisica, Universit\'{a} degli Studi Roma Tre, Via della Vasca Navale 84, I-00146 Roma, Italy}

\author[0000-0003-2686-9241]{D. Stern}
\affiliation{Jet Propulsion Laboratory, California Institute of Technology, Pasadena, CA 91109, USA}

\author[0000-0002-0745-9792]{C.\,M. Urry}
\affiliation{Yale Center for Astronomy \& Astrophysics, 52 Hillhouse Avenue, New Haven, CT 06511, USA}
\affiliation{Department of Physics, Yale University, P.O. Box 2018120, New Haven, CT 06520, USA}

%%%%%%%%%%%%%%%%%%%%%%%%%%%%%%%%%%%%%%%%%%%%%%%%%%%%%%%%%%%%%%%%%%%%%%%%%%%%%%%%%%%%%%%%%%%%%%%
\begin{abstract} %%%%%%%%%%%%%%%%%%%%%%%%%%%%%%%%%%%%%%%%%%%%%%%%%%%%%%%%%%%%%%%%%%%%%%%%%%%%%%
Broadband X-ray spectroscopy of the X-ray emission produced in the coronae of active galactic nuclei (AGN) can provide important insights into the physical conditions very close to their central supermassive black holes. The temperature of the Comptonizing plasma that forms the corona is manifested through a high-energy cutoff that has been difficult to directly constrain even in the brightest AGN because it requires high-quality data at energies above 10 keV. In this paper we present a large collection of coronal cutoff constraints for obscured AGN based on a sample of 130~AGN selected in the hard X-ray band with \swiftbat\ and observed nearly simultaneously with \nustar\ and \swiftxrt. We find that under a reasonable set of assumptions regarding partial constraints the median cutoff is well constrained to $290\pm20$\,keV, where the uncertainty is statistical and given at the 68\,\% confidence level. We investigate the sensitivity of this result to our assumptions and find that consideration of various known systematic uncertainties robustly places the median cutoff between 240\,keV and 340\,keV. The central 68\,\% of the intrinsic cutoff distribution is found to be between about 140\,keV and 500\,keV, with estimated uncertainties of 20\,keV and 100\,keV, respectively. In comparison with the literature, we find no clear evidence that the cutoffs in obscured and unobscured AGN are substantially different. Our analysis highlights the importance of carefully considering partial and potentially degenerate constraints on the coronal high-energy cutoff in AGN.
\end{abstract} %%%%%%%%%%%%%%%%%%%%%%%%%%%%%%%%%%%%%%%%%%%%%%%%%%%%%%%%%%%%%%%%%%%%%%%%%%%%%%%%
%%%%%%%%%%%%%%%%%%%%%%%%%%%%%%%%%%%%%%%%%%%%%%%%%%%%%%%%%%%%%%%%%%%%%%%%%%%%%%%%%%%%%%%%%%%%%%%

%%%%%%%%%%%%%%%%%%%%%%%%%%%%%%%%%%%%%%%%%%%%%%%%%%%%%%%%%%%%%%%%%%%%%%%%%%%%%%%%%%%%%%%%%%%%%%%
\section{Introduction} %%%%%%%%%%%%%%%%%%%%%%%%%%%%%%%%%%%%%%%%%%%%%%%%%%%%%%%%%%%%%%%%%%%%%%%%
\label{sec:intro} %%%%%%%%%%%%%%%%%%%%%%%%%%%%%%%%%%%%%%%%%%%%%%%%%%%%%%%%%%%%%%%%%%%%%%%%%%%%%
%%%%%%%%%%%%%%%%%%%%%%%%%%%%%%%%%%%%%%%%%%%%%%%%%%%%%%%%%%%%%%%%%%%%%%%%%%%%%%%%%%%%%%%%%%%%%%%

The X-ray continuum emission in Active Galactic Nuclei (AGN) is thought to be produced in a region filled with a hot plasma---called the {\em corona}---which up-scatters optical and UV photons into the X-ray band through inverse Compton scattering (e.g., \citealt{vaiana+rosner-1978}, \citealt{haardt+maraschi-1991}, \citealt{merloni+fabian-2003}). Studies of the black hole mass dependence of AGN X-ray variability (e.g., \citealt{axelsson+2013}, \citealt{mcHardy-2013}, \citealt{ludlam+2015}), reverberation of X-ray radiation reprocessed by the accretion disk (e.g., \citealt{deMarco+2013}, \citealt{uttley+2014}, \citealt{kara+2016-lags-summary}), and quasar microlensing (e.g., \citealt{mosquera+2013-oneQSOmicrolens}, \citealt{chartas+2016}, \citealt{guerras+2017}) all suggest that the AGN X-ray source is small in size and located close to the central supermassive black hole (SMBH) and accretion disk, with a typical size scale of 2--20 gravitational radii of the black hole.

Broadband X-ray spectroscopy of the X-ray emission produced in the Comptonizing plasma can provide important insights into the principal properties of the corona, such as its temperature ($kT_e$) and optical depth ($\tau_e$), and ultimately, its geometry. The high-energy cutoff (\ecut) is a parameter of the commonly used approximation to the coronal continuum (e.g., \citealt{rothschild+1983-firstBroadbandXray}, \citealt{gondek+1996-firstBroadbandXray}, \citealt{dadina-2008}): the power law with an exponential cutoff, $\propto E^{-\Gamma}\,e^{-E/E_{\rm \tiny cut}}$, where $E$ is photon energy and $\Gamma$ is the photon index. Spectral parameters of this phenomenological model correspond to the physical parameters of the corona (e.g., \citealt{poutanen+svensson-1996}, \citealt{petrucci+2001}, \citealt{fabian+2015}), and \ecut\ is a proxy for the coronal temperature \citep{middei+2019}. However, it has been difficult to constrain it observationally even in the brightest AGN, since it additionally requires high-quality data above 10\,keV. As a result, our knowledge of the physical properties of the corona remains limited.

Many previous studies of \ecut\ in nearby AGN implied that its typical value is of the order of 100\,keV, starting already with early studies using hard X-ray data from {\em HEAO-1} and {\em CGRO}/OSSE (e.g., \citealt{rothschild+1983-firstBroadbandXray}, \citealt{gondek+1996-firstBroadbandXray}, \citealt{zdziarski+2000}). Coronal cutoffs in unobscured AGN have been the topic of many observational studies over the past decade (e.g., \citealt{panessa+2008-integraltype1}, \citealt{ricci+2011}, \citealt{malizia+2014-type1cutoff}). The field has gained new momentum since the launch of \nustar\ \citep{harrison+2013}, whose focusing ability  boosted data quality in the hard X-ray band (3--79\,keV) far above that of nonfocusing instruments. Summaries of early \nustar\ measurements for individual bright AGN can be found in \citet{fabian+2015} and \citet{tortosa+2018-sample}. \citet{kamraj+2018} and \citet{molina+2019} recently presented \ecut\ constraints for larger samples of unobscured AGN observed with \nustar\ selected using \swiftbat\ and \integral, respectively.

Constraints on \ecut\ in obscured AGN are scarce in the literature despite the prevalence of such AGN in the local universe, and especially at high redshift (e.g., \citealt{buchner+2015}, \citealt{hickox+alexander-2018-obscuredAGNreview}, \citealt{ananna+2019-cxbSynthesis}). Their collective contribution dominates the cosmic X-ray background (CXB) around its broad peak at 30\,keV. The median \ecut\ is an important ingredient of CXB synthesis models (e.g., \citealt{comastri+1995-cxb}, \citealt{gilli+2007}, \citealt{akylas+2012}) that enable us to distinguish different AGN populations and probe their evolution. In particular, at $\gtrsim100$\,keV contributions from unobscured and obscured Seyfert galaxies become comparable to those of blazars (\citealt{ajello+2009-cxbblazars}, \citealt{draper+ballantyne-2009}), with the exact proportions depending directly on the assumed typical \ecut. Through CXB modeling, better constraints on \ecut\ also help to evaluate the importance of relativistic light bending to X-ray spectra of AGN \citep{gandhi+2007-cxb} and the prevalence of highly spinning SMBHs \citep{vasudevan+2016-cxb+spin}.

\ecut\ has been well constrained using \nustar\ data in only a handful of obscured AGN thus far: NGC\,4945 \citep{puccetti+2014-ngc4945}, MCG\,--05-23-016 \citep{balokovic+2015-mcg5}, ESO\,103-G035 \citep{buisson+2018-eso103}, Mrk\,1498 \citep{ursini+2018-obscuredRGs}, NGC\,262, NGC\,2992, NGC\,7172 \citep{rani+2019}, NGC\,2110, and NGC\,4388 \citep{ursini+2019-ngc2110+4388}. Additionally, lower limits on \ecut\ based on \nustar\ data have been derived for obscured sources NGC\,5506 \citep{matt+2015-ngc5506}, Cyg\,A \citep{reynolds+2015-cyga}, Cen\,A \citep{fuerst+2016-cenA}, PKS\,2331--240, and PKS\,2356--61 \citep{ursini+2018-obscuredRGs}. Some constraints based on lower-quality data from nonfocusing hard X-ray instruments exist in the literature, e.g., from \bepposax\ \citep{dadina-2007}, \suzaku\ \citep{tazaki+2011}, \rxte\ \citep{rivers+2011}, \integral\ \citep{deRosa+2012}, \swiftbat\ (\citealt{vasudevan+2013}, \citealt{ricci+2017-bass}), or a combination thereof \citep{molina+2013}.

In this paper we present the first large sample of coronal cutoff constraints for obscured AGN with data quality that exceeds that of non-focusing telescopes in the hard X-ray band. The work presented here is based on a survey of the nearby obscured AGN population with \nustar\ and \swift, sampling over 130~bright AGN selected from the \swiftbat\ all-sky survey (M. Balokovi\'{c} et al., in prep., Paper~I hereafter; also \citealt{balokovic-2017}). While studies of larger samples with more limited data and individual AGN with higher-quality data exist in the literature (e.g., \citealt{ricci+2017-bass}, \citealt{zoghbi+2017-mcg5}, respectively), our study represents an unprecedented combination of excellent data quality and large sample size. This provides a unique opportunity to characterize the average properties of the corona in the highly obscured population and enables comparative studies of coronae in different kinds of AGN.

%%%%%%%%%%%%%%%%%%%%%%%%%%%%%%%%%%%%%%%%%%%%%%%%%%%%%%%%%%%%%%%%%%%%%%%%%%%%%%%%%%%%%%%%%%%%%%%
\section{Data Selection and Analysis} %%%%%%%%%%%%%%%%%%%%%%%%%%%%%%%%%%%%%%%%%%%%%%%%%%%%%%%%%
\label{sec:data} %%%%%%%%%%%%%%%%%%%%%%%%%%%%%%%%%%%%%%%%%%%%%%%%%%%%%%%%%%%%%%%%%%%%%%%%%%%%%%
%%%%%%%%%%%%%%%%%%%%%%%%%%%%%%%%%%%%%%%%%%%%%%%%%%%%%%%%%%%%%%%%%%%%%%%%%%%%%%%%%%%%%%%%%%%%%%%

% ---------------------------------------------------------------------------------------------
\subsection{Short Summary of Paper~I} % -------------------------------------------------------
\label{sec:data-paperone} % -------------------------------------------------------------------
% ---------------------------------------------------------------------------------------------

Data selection, processing, and the bulk of spectral analysis are described in detail in Paper~I; here we only provide a brief summary. The sample used in both works is based on the \swiftbat\ AGN catalog compiled using 70~months of data collection \citep{baumgartner+2013}, which selects AGN bright in the hard X-ray band (14--195\,keV) with high completeness for obscured AGN up to high obscuring columns \citep{ricci+2017-bass}. We focus on a group of AGN optically classified as types~2, 1.9, and 1.8, including some narrow-line LINERs, regardless of their infrared or polarized spectra. Both in Paper~I and here, we refer to this group as type II Seyfert galaxies (Sy\,II), denoted with roman numerals to emphasize the difference with the optical classes differentiated by arabic numerals. The optical class was taken from the \swiftbat\ catalog supplemented by the updated spectroscopy and more uniform classification (following \citealt{osterbrock-1981}) provided by the BAT AGN Spectroscopic Survey (BASS;\footnote{\url{https://www.bass-survey.com/}} \citealt{koss+2017-bass}).

For the majority of our sample, the observed X-ray spectrum was built from a single \nustar\ observation (median exposure 21\,ks), a contemporaneous \swiftxrt\ observation (median exposure 6\,ks), and a time-averaged \swiftbat\ spectrum (effective exposure $\simeq8$\,Ms). Any targets for which \nustar\ or \swift\ data were coadded or excluded are marked with a note in Table~\ref{tab:main}. The median source count rate is 0.1\,ct\,s$^{-1}$ per \nustar\ module (FPM), compared to $2\times10^{-5}$\,ct\,s$^{-1}$ for \swiftbat. For grouping of the \nustar\ and \swiftxrt\ spectra into energy bins, we used a custom procedure described by \citet{balokovic-2017} that results in the total number of bins proportional to data quality and a roughly constant signal-to-noise ratio (SNR) per bin, with a floor at SNR$>\!3$. In spectral fitting, we used \swiftbat\ data simultaneously with \nustar\ and \swiftxrt\ data except in several cases of substantially different flux, as noted in Table~\ref{tab:main}.

The key spectral analysis presented in Paper~I is based on fitting a simple phenomenological spectral model for obscured AGN in \xspec\ \citep{arnaud-1996}. The data selection briefly described above and the spectral analysis define our {{\em uniform sample}}. Numbering 130 AGN, it covers more than 50\,\% of the parent \swiftbat\ 70-month Sy\,II sample and is statistically consistent with a random draw from that sample. It excludes AGN with updated optical classification inconsistent with Sy\,II and AGN with complex X-ray spectra inconsistent with our chosen X-ray spectral model.

The model we employ is used ubiquitously in the literature and consists of components typically observed in X-ray spectra of obscured AGN: an intrinsic cutoff power-law continuum, absorbed by a neutral column density, with reprocessing features represented by a \pexrav\ \citep{magdziarz+zdziarski-1995} continuum and a narrow \feka\ line at 6.4\,keV in the rest frame, and a secondary power law (from extranuclear scattering on free electrons) emerging unabsorbed in the soft X-ray band. In \xspec, the model expression is
\begin{eqnarray*}
    m &=& c_{\mbox{\scriptsize ins}}\!\times\texttt{phabs}\!\times\!(\texttt{zphabs}\!\times\!\texttt{cabs}\!\times\!\texttt{cutoffpl}\\
      &+&f_{\mbox{\scriptsize s}}\!\times\!\texttt{cutoffpl}+\texttt{zgauss}+\texttt{pexrav}),
\end{eqnarray*}
where $c_{\mbox{\scriptsize ins}}$ is an instrumental cross-normalization factor. We refer to this model as the {{\em full model}}. Its free spectral parameters are the photon index ($\Gamma$), line--of-sight column density (\nh), Compton hump normalization (\rpex, the absolute value of the negative $R$ parameter in \pexrav), \feka\ equivalent width (\feew), and the relative normalization of the secondary power law (\fsca). Paper~I presents constraints on these phenomenological spectral parameters for all 130 AGN in the {{\em uniform sample}} while keeping \ecut\ fixed at 300\,keV.

\begin{figure}
\begin{center}
\vspace{0.3cm}
\includegraphics[width=1.0\columnwidth]{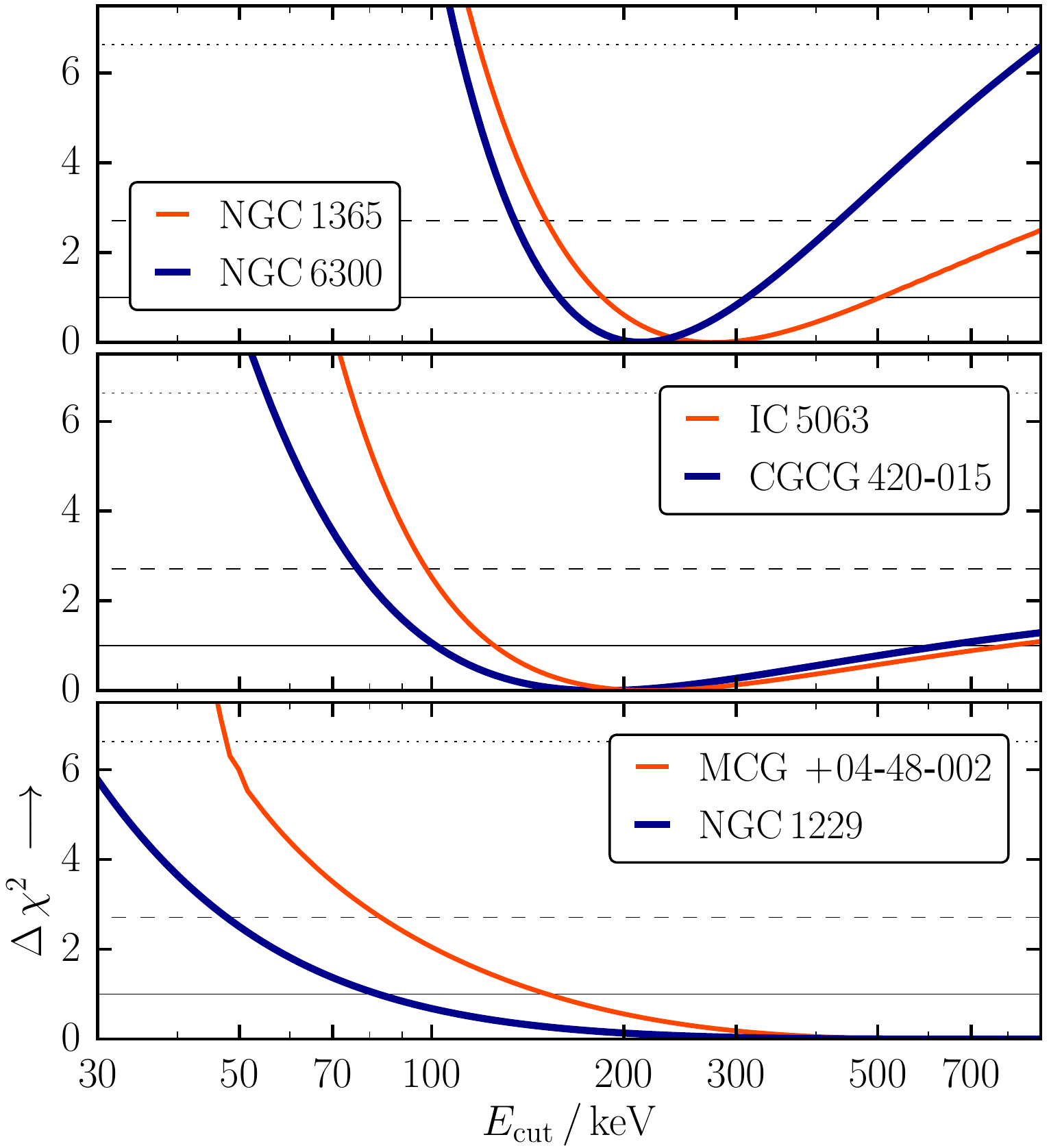}
\caption{ Representative examples of $\Delta \chi^2$ curves over the \ecut\ parameter space. Horizontal solid, dashed, and dotted lines mark the $\Delta \chi^2$ levels corresponding to 68, 90, and 99 per cent confidence, respectively. All other spectral parameters are free to vary. In the top panel \ecut\ is fully constrained (yielding a best-fit value and two-sided uncertainty), in the middle panel it is partially constrained (lower limit and a weakly constrained best-fit value), and in the bottom panel examples provide only a lower limit.
\label{fig:chitwo}}
\vspace{0.2cm}
\end{center}
\end{figure}

\begin{figure*}[t!]
\begin{center}
\includegraphics[width=1.0\textwidth]{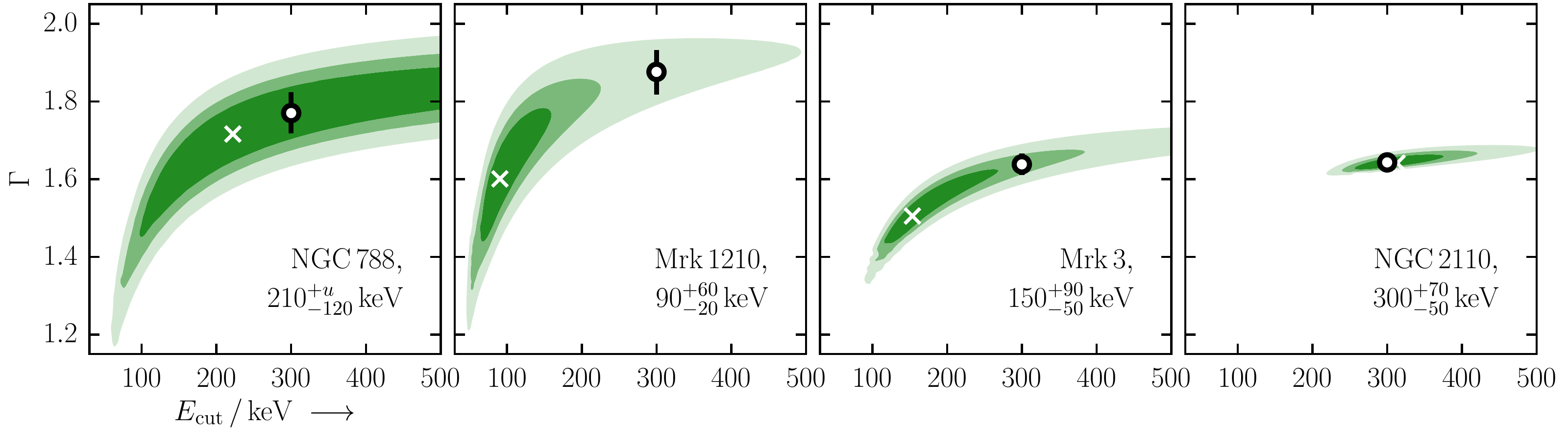}
\caption{ Three representative examples of the degeneracy between parameters $\Gamma$ and \ecut\ shown as 1\,$\sigma$, 2\,$\sigma$, and 3\,$\sigma$ contours (going from darker to lighter green) in the parameter space spanned by these two spectral parameters. In each panel, the best fit is marked with a white cross, and the open black circle with an error bar shows the best-fit $\Gamma$ and its 1\,$\sigma$ uncertainty with the assumption of \ecut\,$=300$\,keV (from a statistically acceptable model presented in Paper~I). From left to right, the figure shows a progression from lowest to highest quality of \ecut\ constraint, with the second panel (Mrk\,1210) showing an example of a result excluded to mitigate the parameter degeneracy (see \S\,\ref{sec:data-degenerate}).
\label{fig:contours}}
\vspace{0.2cm}
\end{center}
\end{figure*}

% ---------------------------------------------------------------------------------------------
\subsection{Spectral Analysis with Free \ecut} % ----------------------------------------------
\label{sec:data-addingecut} % -----------------------------------------------------------------
% ---------------------------------------------------------------------------------------------

The study presented here continues the analysis presented in Paper~I by additionally letting the \ecut\ parameter be fitted instead of being fixed at 300\,keV. This does not typically result in a significantly better fit to our data, but does provide previously unavailable information on obscured AGN coronae. Data for three AGN in our sample (Cen\,A, NGC\,5506, and MCG\,--05-23-016) require a fitted \ecut\ in order to reach a statistically acceptable $\chi^2$ according to our adopted null-probability (\pnull) rejection threshold at 5\,\%. Spectral models with free \ecut\ are therefore already discussed in Paper~I for these targets. For the rest of the sample we simply include \ecut\ as an additional free parameter in order to test whether the data yield at least partial constraints on \ecut\ from spectral fitting. Since \ecut\ is an additional parameter in a model that already fits the data for all sources well, the resulting improvements in terms of $\chi^2$ are not statistically significant in most cases. As expected, the obtained constraints are consistent with 300\,keV in nearly all cases, although not always within the derived 68\,\% confidence interval. Throughout this paper, as well as in Paper~I, we define the single-parameter 68\,\% confidence interval using a difference of $\Delta \chi^2=1$ with respect to the best fit while allowing the other parameters to freely vary in the fit.

Spectral analysis for the whole sample presented in Paper~I resulted in constraints on \ecut\ for 114 AGN in total. There is a clear difference in quality of constraints: some spectra yield the best-fit value, $E_{\rm \small cut}^{\rm \small \,b.f.}$, with both lower and upper limits of the 68\,\% confidence interval ($E_{\rm \small cut}^{\rm \small \,l.l.}$ and $E_{\rm \small cut}^{\rm \small \,u.l.}$, respectively), while others yield only partial constraints. Examples of three constraint classes are shown in Figure~\ref{fig:chitwo}. Out of 130 AGN considered in Paper~1, we find full constraints for 45, partial constraints for 69 (of which 50 yield only $E_{\rm \small cut}^{\rm \small \,l.l.}$, while for 19 we also find a $E_{\rm \small cut}^{\rm \small \,b.f.}$), and no constraints for 16. Although \xspec\ formally finds $E_{\rm \small cut}^{\rm \small \,b.f.}$ for more than 64 targets, in the remainder of such cases the difference in $\chi^2$ from the best-fit value ($\Delta \chi^2$) for \ecut\,$>\,E_{\rm \small cut}^{\rm \small \,b.f.}$ is so small that we cannot consider $E_{\rm \small cut}^{\rm \small \,b.f.}$ reliably constrained.\footnote{In simulated data with the same spectral parameters and matching data quality, these values cannot be consistently recovered. Although $\Delta \chi^2 >1$ (as in the examples shown in the middle panel of Figure~\ref{fig:chitwo}) formally implies that $E_{\rm \small cut}^{\rm \small \,u.l.}$ can be defined, simulations we performed strongly suggest that our data cannot reliably constrain these values.} Likewise, fits with $E_{\rm \small cut}^{\rm \small \,u.l.}>500$\,keV are unlikely to be reliable except in a few cases of high-quality data. We discard such $E_{\rm \small cut}^{\rm \small \,b.f.}$ and $E_{\rm \small cut}^{\rm \small \,u.l.}$ values on the basis of limited data quality. The statistics of constraints quoted above already include these quality-based cuts. The spectral fitting results described in this section are listed in Table~\ref{tab:main}.

% ---------------------------------------------------------------------------------------------
\subsection{Exclusion of Potentially Degenerate Constraints} % --------------------------------
\label{sec:data-degenerate} % -----------------------------------------------------------------
% ---------------------------------------------------------------------------------------------

Spectral parameters of the model defined in \S\,\ref{sec:data-paperone} are not fully independent, leading to degeneracy between fitted parameters that depends on both data quality and the spectral shape. The degeneracy between parameters $\Gamma$ and \ecut is well known in the literature, as it also affects studies of \ecut\ in unobscured AGN (e.g., \citealt{tortosa+2018-twoagn}, \citealt{kamraj+2018}, \citealt{molina+2019}). In Figure~\ref{fig:contours} we provide several examples. Unlike other pairs of parameters in our model, $\Gamma$ and \ecut\ are always tightly correlated in the same direction, leading to a cumulative, systematic effect in studies of large samples. For obscured AGN, this problem is exacerbated, as obscuration limits the energy range over which the coronal continuum is sampled directly.

In fitting spectra over a limited energy band, not all constraints on \ecut\ should be considered credible. Some very low best-fit values of \ecut\ are a simple consequence of a simultaneously low \ecut\ and $\Gamma$. Although the spectral shape formally fits the data very well, such combinations of $\Gamma$ and \ecut\ do not correspond to a spectrum that could realistically be generated in the physical conditions expected in an AGN corona (see, e.g., \citealt{zdziarski+lightman-1985}, \citealt{stern+1995}, \citealt{poutanen+svensson-1996}). In order to avoid a possible bias from spurious low-\ecut\ and low-$\Gamma$ fits, we devised a simple way to separate credible \ecut\ constraints from those potentially affected by the parameter degeneracy via a cut in implied electron scattering optical depth ($\tau_e$), which can be approximated as a function of $\Gamma$ and \ecut\ under reasonable assumptions.

For the conversion of phenomenological spectral parameters $\Gamma$ and \ecut\ to basic physical parameters of the corona, $kT_e$ and $\tau_e$, we adopt the following approximations. First, $kT_e=$\,\ecut$/j(\tau_e)$, where
\begin{equation}
\label{eq:kt}
  j (\tau_e) =
  \begin{cases}
    \ 2, & \mbox{if}\ \tau_e \leq 1, \\
    \ \tau_e + 1, & \mbox{if}\ 1<\tau_e<2, \\
    \ 3, & \mbox{if}\ \tau_e \geq 2.
  \end{cases}
\end{equation}
We construct the continuous relation above based on two commonly assumed limiting cases for $kT_e$, \ecut$/2$ for $\tau_e \leq1$ and \ecut$/3$ for $\tau_e \gg1$ (e.g., \citealt{shapiro+1976}), noting that more realistic coronal models cover a wider range of scaling factors \citep{middei+2019}. $kT_e$, $\Gamma$, and $\tau_e$ are related via an approximate expression derived for a plane-parallel corona and formally valid for $\tau_e\gtrsim1$ (\citealt{zdziarski-1985}, \citealt{petrucci+2001}):
\begin{equation}
   \Gamma = \sqrt{\,\frac{9}{4}+\frac{511\,\mbox{\rm \small keV}}{kT_e \tau_e (1+\tau_e/3)}} - \frac{1}{2}
\end{equation}
Inverting this relation, we compute $\tau_e$ from measured $\Gamma$ and \ecut\ by iteratively solving the following equation:
\begin{equation}
\label{eq:tauline}
   \tau_e (1+\tau_e/3) / j(\tau_e) = \frac{511\,\mbox{\rm \small keV}/E_{\rm \small cut}}{{\left( \Gamma + 1/2 \right)}^2-9/4}
\end{equation}

Equation~\ref{eq:tauline} defines a series of curves with constant $\tau_e$ in the plane spanned by parameters \ecut\ and $\Gamma$, as shown in Figure~\ref{fig:taucutoff}. For our fiducial analysis, we choose to separate credible \ecut\ constraints from the likely degenerate ones with a $\tau_e\!<\!3$ cut. This cut should not be interpreted as a physical limit on the optical depth; rather, it is a practical way of excluding potentially degenerate constraints. While the exact value for the cut is arbitrary, we argue that one has to be made in order to avoid bias that may be purely due to the \ecut--$\Gamma$ degeneracy. This is examined in more detail in \S\,\ref{sec:discussion}.

%%%%%%%%%%%%%%%%%%%%%%%%%%%%%%%%%%%%%%%%%%%%%%%%%%%%%%%%%%%%%%%%%%%%%%%%%%%%%%%%%%%%%%%%%%%%%%%
\section{Results} %%%%%%%%%%%%%%%%%%%%%%%%%%%%%%%%%%%%%%%%%%%%%%%%%%%%%%%%%%%%%%%%%%%%%%%%%%%%%
\label{sec:results} %%%%%%%%%%%%%%%%%%%%%%%%%%%%%%%%%%%%%%%%%%%%%%%%%%%%%%%%%%%%%%%%%%%%%%%%%%%
%%%%%%%%%%%%%%%%%%%%%%%%%%%%%%%%%%%%%%%%%%%%%%%%%%%%%%%%%%%%%%%%%%%%%%%%%%%%%%%%%%%%%%%%%%%%%%%

% ---------------------------------------------------------------------------------------------
\subsection{Summary of Individual Constraints} % ----------------------------------------------
\label{sec:results-individual} % --------------------------------------------------------------
% ---------------------------------------------------------------------------------------------

Our sample covers a factor of $\simeq\!50$ in hard X-ray flux. \nustar\ exposures (3.7--57\,ks) do not scale with the targets' flux, so the number of degrees of freedom ($\nu$, a proxy of overall data quality) ranges between 30 and~770. Targets with $\nu\!\lesssim\!60$ consistently do not provide any \ecut\ constraints. For $\nu\!\gtrsim\!60$ the data typically provide lower limits on \ecut, and full constraints in some cases. Higher-quality data generally allow for stronger, more stringent constraints with a range demonstrated in Figure~\ref{fig:contours}. The relative uncertainty in \ecut\ is not just a simple function of~$\nu$. The smallest relative uncertainties are found for AGN with constraints in the discarded $\tau_e\!>\!3$ group, which contains targets over the full range of data quality, in particular those with high obscuring columns (\nh\,$>\!10^{24}$\,\cmmt; e.g., NGC\,3281, NGC\,6240) or prominent Compton humps (\rpex\,$>\!1$; e.g., Mrk\,1210, NGC\,1194). Additional curvature due to these features makes spectra more susceptible to the \ecut--$\Gamma$ degeneracy.

\begin{figure}[t]
\begin{center}
\includegraphics[width=1.0\columnwidth]{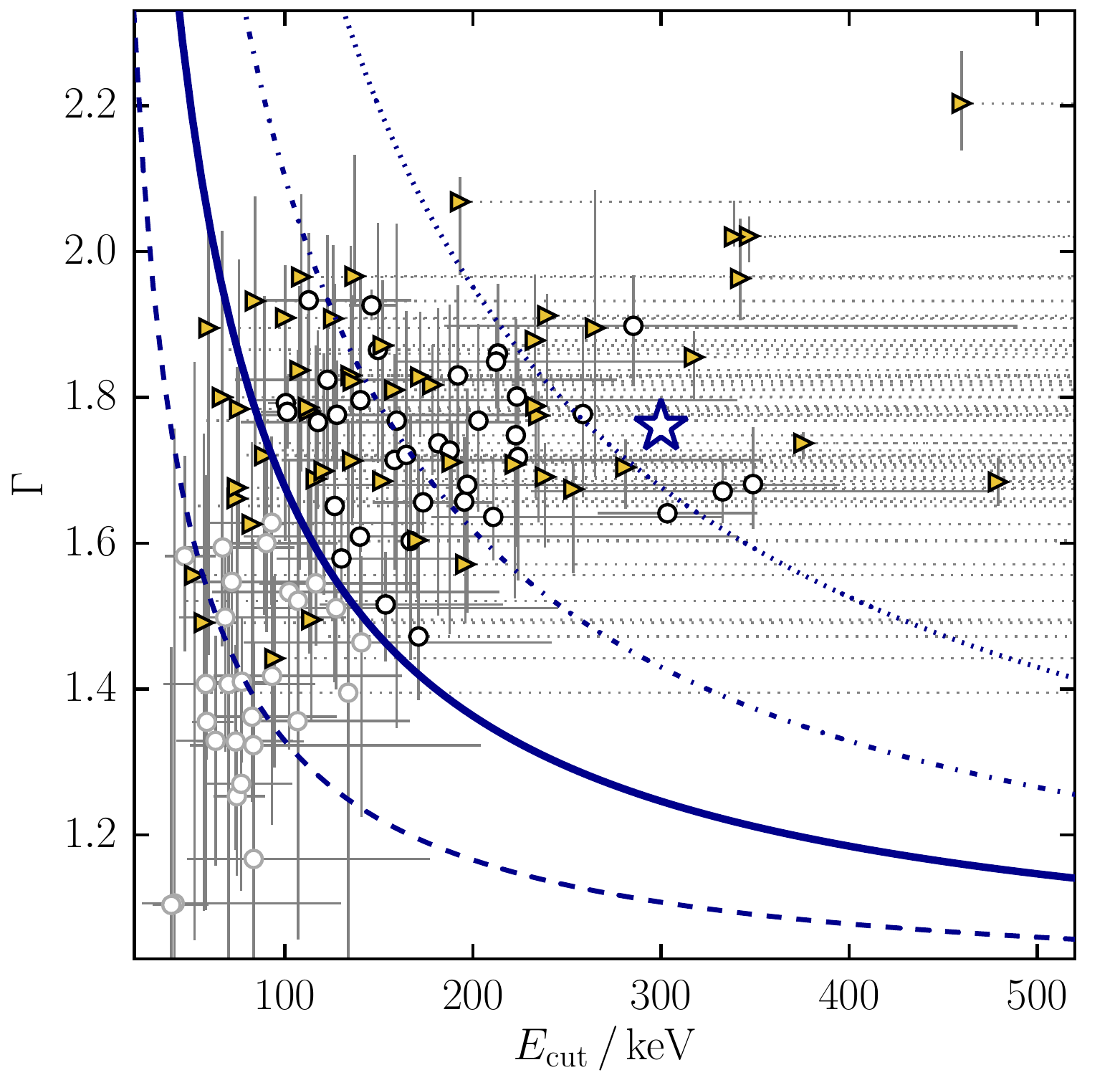}
\caption{ Distribution of spectral fitting results in the \ecut--$\Gamma$ plane. White-filled circles mark best-fit values where they could be obtained, while yellow-filled triangles mark lower limits on \ecut. Thin gray lines show individual error bars, with those that formally extend above 500\,keV shown with dotted lines for clarity. The blue star marks the median $\Gamma$ obtained with fixed \ecut\,$=300$\,keV in Paper~I. The blue lines are defined by Equation~\ref{eq:tauline} and mark $\tau_e=1, 2, 3, \mbox{and}\ 5$ with dotted, dash-dotted, solid, and dashed lines, respectively. We discard constraints with a best-fit value below the solid line in order to minimize bias due to the \ecut--$\Gamma$ degeneracy which causes some fits to drift to extremely low values of both \ecut\ and $\Gamma$ (see Figure~\ref{fig:contours}).
\label{fig:taucutoff}}
\vspace{0.2cm}
\end{center}
\end{figure}

Treating \ecut\ as a free parameter in spectral fitting results typically lowers the best-fit $\chi^2$ compared to fixed \ecut\,$=300$\,keV. However, in most cases, the reduction in $\chi^2$ is not statistically significant because of the limited data quality and the fact that our analysis starts from a model that already fits the data well. Even in some cases of high-quality data, treating \ecut\ as a free parameter does not result in a significantly better fit because the uncertainty interval is consistent with 300\,keV; such examples are NGC\,2110 ($300_{-40}^{+50}$\,keV) and NGC\,4388 ($210_{-40}^{+120}$\,keV). The lack of significant decreases in $\chi^2$ simply justifies \ecut\,$=300$\,keV as a good assumption for the spectral analysis presented in Paper~I. The majority of spectra that formally do not include 300\,keV within the derived 68\,\% uncertainty intervals belong to the discarded $\tau_e\!>\!3$ group: 25 out of 44 in total. Of the 87 constraints with $\tau_e\!<\!3$, 10 are above 300\,keV, 9 are below, and 68 are consistent with this value given their 1\,$\sigma$ uncertainties derived from \xspec.

The typical \ecut\ constraint afforded by the spectral analysis of our sample is a lower limit in the 100--300\,keV range. Excluding constraints with $E_{\rm \small cut}^{\rm \small \,b.f.}$ corresponding to $\tau_e\!>\!3$ (the majority of which are full constraints) and the top third of our sample by data quality (accounting for 15 out of the remaining 20 full constraints), the rest of the sample yields \ecut\ constraints qualitatively similar to those shown in the middle and bottom panels of Figure~\ref{fig:chitwo}. Stepping through the \ecut\ parameter space, $\chi^2$ typically increases toward the lower end, and in some cases increases by a small amount above the best-fit value found in \xspec. Our chosen confidence level (68\,\%, corresponding to $\Delta \chi^2=1$) implies that typical constraints for any individual AGN should be considered with due caution. We note that despite the relatively low confidence level, for sample statistics such constraints are more informative than a smaller number of higher-confidence limits in the range of 30--100\,keV.

Our sample also includes some strong constraints for bright AGN. Three targets in our sample require \ecut\ other than 300\,keV to reach $\chi^2$ low enough to even consider the spectral model statistically acceptable based on the \pnull\,$>\!5$\,\% criterion: MCG\,--05-023-16, NGC\,5506, and Cen\,A. For a further three, NGC\,262, NGC\,7172, and ESO\,103-G035, treating \ecut\ as a free parameter resulted in a significantly better fit. A notable improvement in best-fit $\chi^2$ with free \ecut\ happens only in the cases where the spectrum is clearly more curved or clearly less curved than under the assumption of \ecut\,$=300$\,keV. For four out of the six AGN mentioned above, we find well-constrained \ecut\ in the 100--200\,keV range. For the remaining two, fits imply \ecut\,$=550_{-90}^{+140}$\,keV (Cen\,A), and \ecut\,$>670$\,keV (NGC\,7172). More details on these extreme cases are given in Appendix~\ref{sec:appendix}.

% ---------------------------------------------------------------------------------------------
\subsection{Distribution of \ecut\ in the Sample} % -------------------------------------------
\label{sec:results-ecutdist} % ----------------------------------------------------------------
% ---------------------------------------------------------------------------------------------

The full distribution of constraints in the \ecut--$\Gamma$ plane obtained through our spectral analysis is shown in Figure~\ref{fig:taucutoff}. A straightforward comparison with Figure~\ref{fig:contours} illustrates the direction of the degeneracy between these two parameters. After exclusion of \ecut\ constraints with $E_{\rm \small cut}^{\rm \small \,b.f.}$ corresponding to $\tau_e\!>\!3$, which is shown with the solid blue line in Figure~\ref{fig:taucutoff}, our data set consists of 87 constraints in total. Of these, 37 include $E_{\rm \small cut}^{\rm \small \,b.f.}$, of which 20 are full constraints. For 50 cases we found only lower limits ($E_{\rm \small cut}^{\rm \small \,l.l.}$; at 68\,\% confidence). This includes 11 constraints with only $E_{\rm \small cut}^{\rm \small \,l.l.}$ that formally fall in the $\tau_e\!>\!3$ regime; as they correspond to upper limits on $\tau_e$, they imply that $\tau_e$ in these cases is most likely lower.

\begin{figure}[t!]
\begin{center}
\vspace{0.2cm}
\includegraphics[width=1.0\columnwidth]{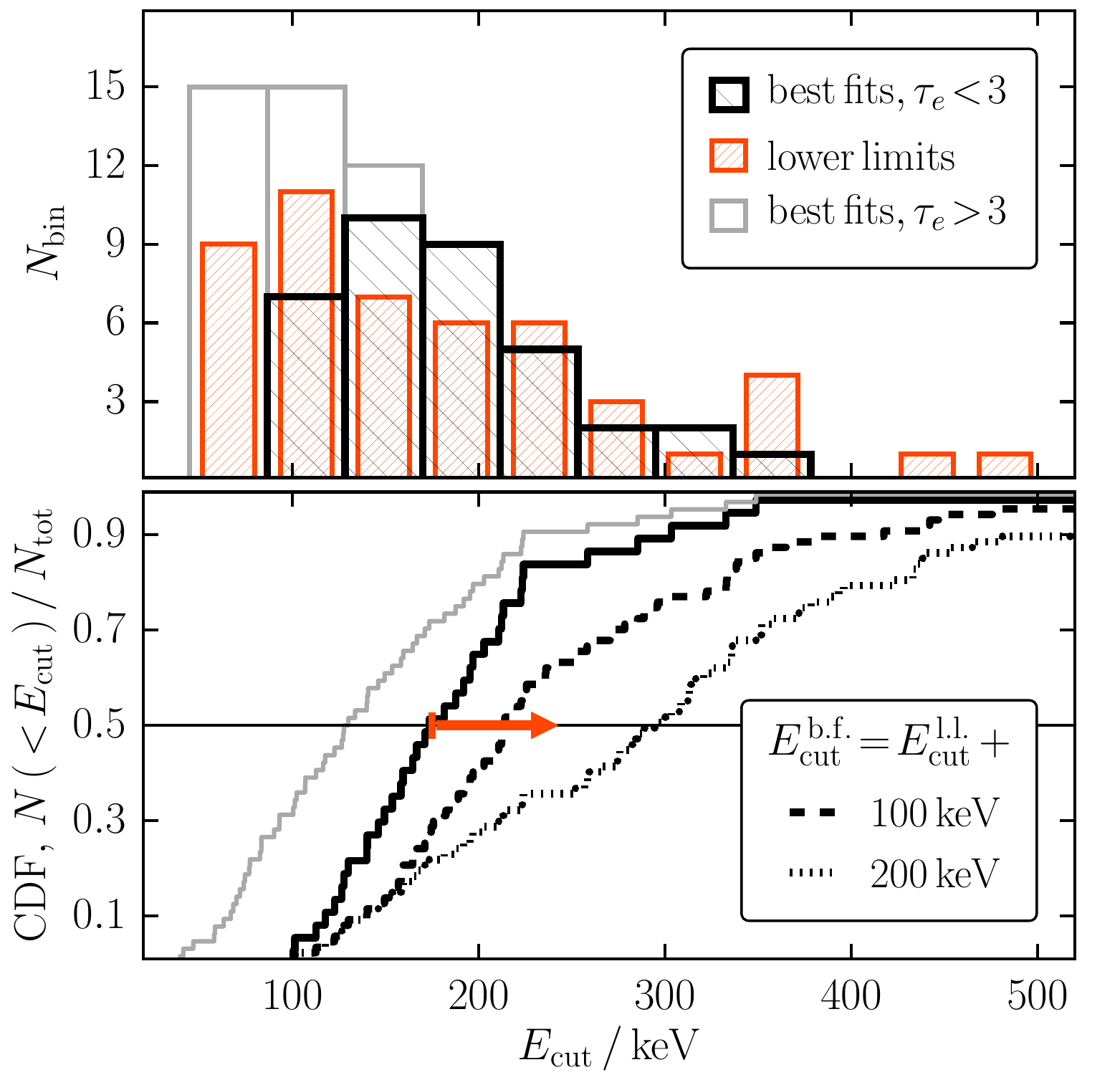}
\caption{ Top panel: histogram of the derived \ecut\ constraints for our sample. Best-fit values ($E_{\rm \small cut}^{\rm \small \,b.f.}$) are shown in black or gray, depending on $\tau_e$ calculated using Equation~\ref{eq:tauline}, while lower limits ($E_{\rm \small cut}^{\rm \small \,l.l.}$) are shown in red. Bottom panel: Cumulative distribution functions (CDFs) based only on $E_{\rm \small cut}^{\rm \small \,b.f.}$ including $\tau_e\!>\!3$ constraints (gray) and excluding them (black). Due to numerous lower limits, these are actually lower bounds on the true CDF and its median (illustrated by the red rightward-pointing arrow). For comparison, we also show CDFs for the complete set of constraints under the simple assumptions that non-constrained $E_{\rm \small cut}^{\rm \small \,b.f.}$ are 100\,keV and 200\,keV above the measured $E_{\rm \small cut}^{\rm \small \,l.l.}$ (dashed and dotted black lines, respectively).
\label{fig:ecutdist}}
\vspace{0.2cm}
\end{center}
\end{figure}

In the top panel of Figure~\ref{fig:ecutdist} we show a histogram of best-fit \ecut\ values ($E_{\rm \small cut}^{\rm \small \,b.f.}$) and lower limits ($E_{\rm \small cut}^{\rm \small \,l.l.}$) in cases where best-fit values are absent. Considering only best-fit values and neglecting any lower limits, the median of the \ecut\ distribution is 180\,keV, with 68\,\% of the distribution covering the range of 100--400\,keV. Including all $E_{\rm \small cut}^{\rm \small \,b.f.}$ derived from spectral fitting, i.e. temporarily neglecting the $\tau_e\!<\!3$ cut described in \S\,\ref{sec:data-degenerate}, would imply that the \ecut\ distribution median is 130\,keV. However, these are clearly biased values failing to account for a considerable number of lower limits that suggest a higher median \ecut. This is illustrated in the bottom panel of Figure~\ref{fig:ecutdist}.

\begin{figure}[t]
\begin{center}
\vspace{0.2cm}
\includegraphics[width=1.0\columnwidth]{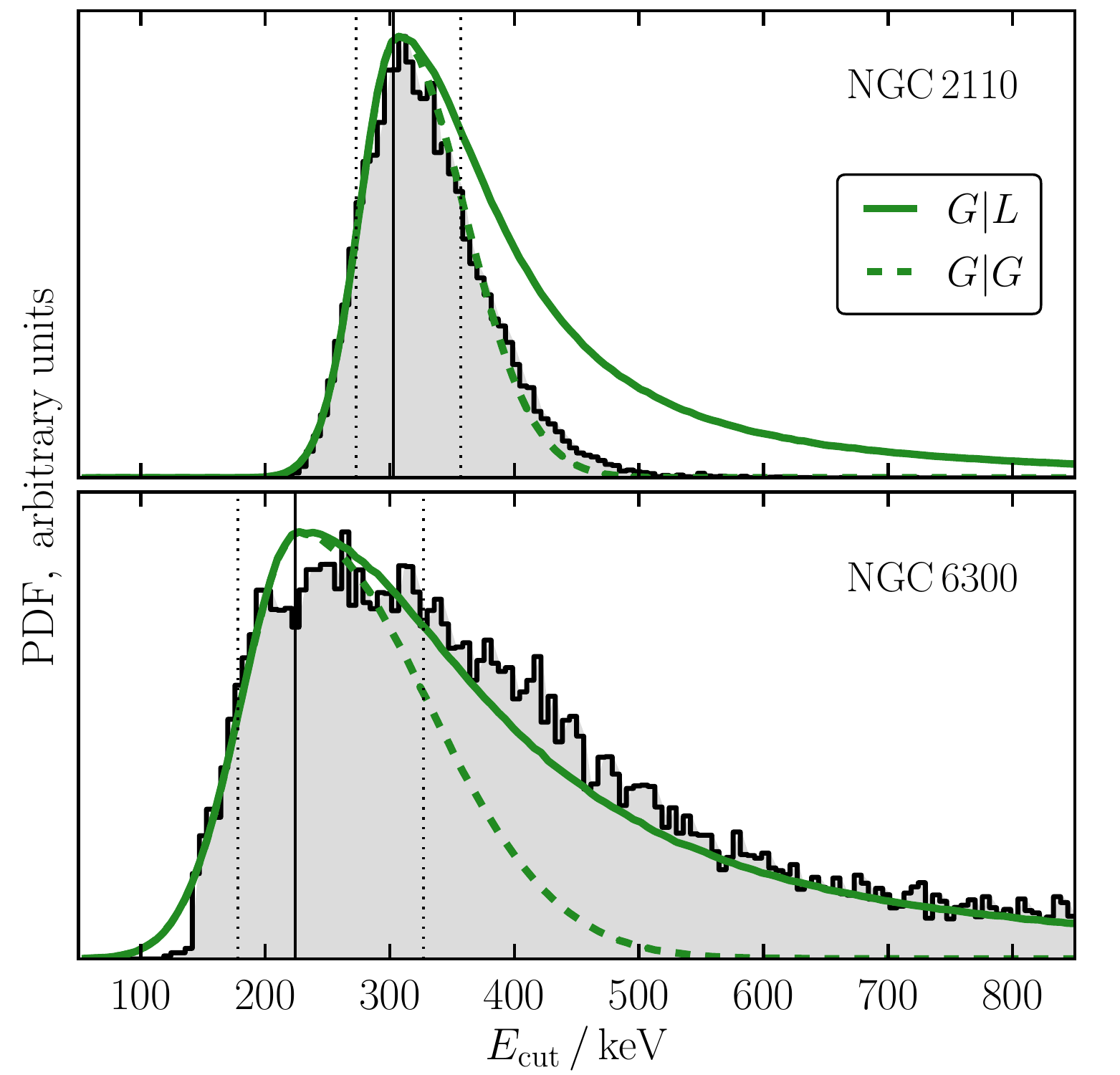}
\caption{ Two illustrative examples of probability density functions (PDFs) derived from MCMC chains for two targets (NGC\,2110 in the top panel, and NGC\,6300 in the bottom panel), shown as gray histograms, compared to assumed analytic PDFs shown with green curves. Solid lines show the adopted fiducial PDFs composed of a Gaussian and a Lorentzian ($G|L$), while the dashed lines show a two-sided Gaussian ($G|G$). In both cases, the transition occurs at the best-fit value ($E_{\rm \small cut}^{\rm \small \,b.f.}$; vertical solid line) and the standard deviation on either side is determined by the extent of the 68\,\% confidence interval ($E_{\rm \small cut}^{\rm \small \,l.l.}$ and $E_{\rm \small cut}^{\rm \small \,u.l.}$; vertical dotted lines).
\label{fig:pdfs}}
\vspace{0.2cm}
\end{center}
\end{figure}

In order to incorporate numerous partial constraints into our estimate of the \ecut\ distribution, we employ a Monte Carlo bootstrapping method (e.g., \citealt{andrae-2010}). We calculate a large number of cumulative distribution functions (CDFs) by resampling each \ecut\ constraint according to an approximate probability density function (PDF) that peaks at the best-fit value and contains 68\,\% of the total probability within the 68\,\% confidence interval determined from \xspec. For lower limits, we assume a particular PDF that contains 16\,\% of the probability (as a one-sided tail outside of the central 68\,\%) below $E_{\rm \small cut}^{\rm \small \,l.l.}$ determined from \xspec. From a collection of 10,000 CDFs constructed in this way, we calculate the median CDF which provides a more informed estimate of the median \ecut\ and the range containing 68\,\% of the \ecut\ distribution. This procedure automatically provides a measure of uncertainty in these numbers based on the spread of the resampled CDFs. As with \ecut\ constraints, we quote uncertainties at the 1\,$\sigma$ level, and round them to the typical accuracy floor of 10\,keV.

Our method requires an assumption of a PDF for resampling the constraints derived from spectral fitting. Taking into account the asymmetry of the likely PDFs for \ecut\ (demonstrated by the representative $\chi^2$ curves shown in Figure~\ref{fig:chitwo}), we adopt a PDF that consists of a Gaussian below $E_{\rm \small cut}^{\rm \small \,b.f.}$ and a Lorentzian above it. The latter provides a notably slower fall-off toward high energies. Both sides are scaled so that $E_{\rm \small cut}^{\rm \small \,b.f.}$ is the peak of the PDF and its width on each side matches the boundaries of the 68\,\% uncertainty interval.\footnote{The Lorentzian formally has undefined moments, but its quantiles can be calculated from its CDF. The equivalent of 1\,$\sigma$ corresponds to $1.817\times\gamma$, where $\gamma$ is the scale parameter of the distribution.} This choice is further supported by visual comparison with two PDFs determined from converged Markov Chain Monte Carlo (MCMC) chains shown in Figure~\ref{fig:pdfs}. A two-sided Gaussian ($G|G$) is an excellent match in the well-constrained PDF for NGC\,2110. However, including a Lorentzian tail to high energies ($G|L$) similar to the PDF for NGC\,6300 is likely more appropriate for our sample given the generally lower reliability of $E_{\rm \small cut}^{\rm \small \,u.l.}$ determinations.

Lower limits are clearly less informative constraints, so we adopt a uniform distribution spanning a range of 300\,keV in total. Its minimum is set so that 16\,\% of probability is below the $E_{\rm \small cut}^{\rm \small \,l.l.}$ determined from spectral fitting. We treat partial constraints with both $E_{\rm \small cut}^{\rm \small \,l.l.}$ and $E_{\rm \small cut}^{\rm \small \,b.f.}$ values (but no upper limit on the confidence interval) as full constraints by assuming that $E_{\rm \small cut}^{\rm \small \,u.l.} = E_{\rm \small cut}^{\rm \small \,b.f.}\!+150$\,keV. The widths of these distributions are set conservatively based on widths of 68\,\% confidence intervals for full constraints: the median width of ($E_{\rm \small cut}^{\rm \small \,u.l.}-E_{\rm \small cut}^{\rm \small \,l.l.}$) is 150\,keV, while the maximum in our sample is approximately 300\,keV. In \S\,\ref{sec:discussion-sysunc} we discuss the effect of these choices on our results and consider alternatives.

The median of the \ecut\ distribution derived for our sample with the procedure and choices described above is $290\pm20$\,keV. This uncertainty refers only to the median of the distribution and reflects only statistical uncertainty. The distribution itself is significantly wider, with 68\,\% of the probability distribution spanning between $140\pm10$\,keV and $540\pm60$\,keV. The lower end of the 68\,\% probability interval can be determined robustly, as it is largely independent of assumptions regarding partial constraints. The shape of the distribution, its median, and especially the upper end of the 68\,\% probability interval depend to a certain extent on assumptions. For the remainder of the paper we adopt the sample and the analysis described here as fiducial. In the following sections we justify particular choices regarding our method, quantify known systematic uncertainties, and examine the effects of alternatives on the inferred \ecut\ distribution.

%%%%%%%%%%%%%%%%%%%%%%%%%%%%%%%%%%%%%%%%%%%%%%%%%%%%%%%%%%%%%%%%%%%%%%%%%%%%%%%%%%%%%%%%%%%%%%%
\section{Discussion} %%%%%%%%%%%%%%%%%%%%%%%%%%%%%%%%%%%%%%%%%%%%%%%%%%%%%%%%%%%%%%%%%%%%%%%%%%
\label{sec:discussion} %%%%%%%%%%%%%%%%%%%%%%%%%%%%%%%%%%%%%%%%%%%%%%%%%%%%%%%%%%%%%%%%%%%%%%%%
%%%%%%%%%%%%%%%%%%%%%%%%%%%%%%%%%%%%%%%%%%%%%%%%%%%%%%%%%%%%%%%%%%%%%%%%%%%%%%%%%%%%%%%%%%%%%%%

% ---------------------------------------------------------------------------------------------
\subsection{Setting Expectations with Simulated Data} % -----------------------------
\label{sec:discussion-sims} % -----------------------------------------------------------------
% ---------------------------------------------------------------------------------------------

\begin{figure}[t]
\begin{center}
\vspace{0.2cm}
\includegraphics[width=1.0\columnwidth]{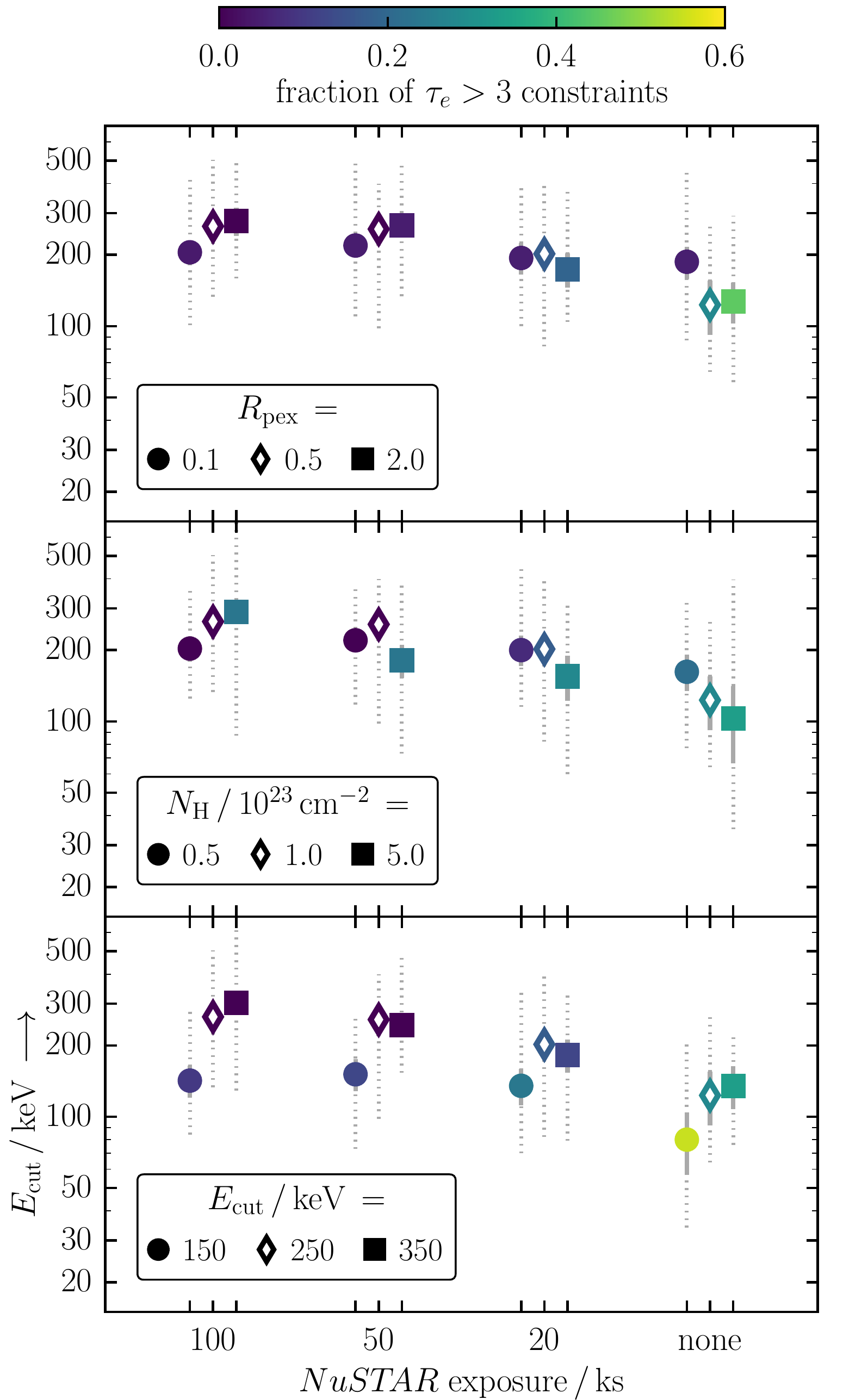}
\caption{ Results of spectral analysis on simulated data showing how the lack of \nustar\ data, as well as high values of \rpex\ and \nh, influence the estimate of \ecut. Solid error bars show the uncertainty in the median value based on 100 simulated measurements, while dotted error bars span the 68\,\% probability interval for each sample. In each panel, the spectral parameter specified in the legend is varied as input for simulated data, which are then fitted to obtain \ecut\ constraints. For the reference model shown with open diamonds in each panel, we assumed the following parameters: ($\Gamma$, \rpex, \nh\,/\,\cmmt, \ecut\,/\,keV)\,$=\,(1.8, 0.5, 1.0\!\times\!10^{23}, 250)$. See \S\,\ref{sec:discussion-sims} for a description of these simulations and a discussion of their implications.
\label{fig:dataqsim}}
\vspace{0.2cm}
\end{center}
\end{figure}

In order to better understand the results and justify some of the choices made for our fiducial analysis, we resort to simulations in which we control the input parameters. We simulate the whole measurement process under different assumptions following the procedure described below. The results are summarized in Figure~\ref{fig:dataqsim}.

For a fiducial model we adopted spectral parameters representative of our sample: ($\Gamma$, \rpex, \nh\,/\,\cmmt, \ecut\,/\,keV)\,$=\,(1.8, 0.5, 1.0\!\times\!10^{23}, 250)$. We set the flux to be representative of the sample median. In each realization, we simulated XRT and BAT spectra with typical exposures of 6\,ks and 10\,Ms, respectively, and \nustar\ spectra (both FPMA and FPMB) with exposures of 20, 50, and 100\,ks. We then performed the same spectral analysis as on real data and collected \ecut\ constraints for each of the \nustar\ exposure lengths, as well as without any \nustar\ data (i.e., \swift\ data only). We repeated this procedure 100 times to approximately match our sample size. We then changed the value of one spectral parameter at a time, repeating the procedure for \rpex\ values 0.1 and 2.0, \nh\ of $5\!\times\!10^{22}$\,\cmmt\ and $5\!\times\!10^{23}$\,\cmmt, and \ecut\ set to 150\,keV and 350\,keV.

The simulations demonstrate the degree of systematic uncertainty in \ecut\ measurements for obscured AGN. Almost independently of the amount of \nustar\ exposure and true underlying parameters of AGN spectra, the 68\,\% spread in individual \ecut\ constraints is always large, typically $\gtrsim\!100$\,keV. This can be understood as a consequence of limited photon statistics in the highest-energy bins (as well as the highest-energy \swiftbat\ channels), since the measurement critically depends on subtle curvature at the high-energy end of the \nustar\ band. Any two measurements for the same input spectrum can differ significantly, so a single constraint should always be considered with caution. Inclusion of higher-quality \nustar\ data (or, equivalently, brighter targets) results in \ecut\ constraints that are both more accurate and more precise. However, even with long \nustar\ exposures on an AGN with our fiducial model spectrum (\ecut\,$=250$\,keV) and flux typical for our sample, the best-fit \ecut\ is expected to appear below 150\,keV and above 450\,keV in roughly 20\,\% of individual measurements.

For the sample medians shown in Figure~\ref{fig:dataqsim} we did not discard $\tau_e\!>3$ constraints so as to demonstrate their impact. Results of our simulations reveal several trends highlighted in Figure~\ref{fig:dataqsim}:
\begin{itemize}
  \vspace{-0.2cm}
  \item The fraction of spuriously inaccurate \ecut\ constraints corresponding to $\tau_e\!>3$ is generally inversely proportional to the \nustar\ exposure.
  \vspace{-0.2cm}
  \item A more pronounced Compton hump (greater \rpex\ parameter) tends to result in a higher fraction of $\tau_e\!>3$ constraints, making the median \ecut\ biased toward low values. The effect decreases with increasing \nustar\ data quality.
  \vspace{-0.2cm}
  \item High obscuration also makes \ecut\ constraints biased toward low values. Almost independently of \nustar\ data quality, a higher column density leads to more scatter in \ecut\ constraints and a higher proportion of $\tau_e\!>3$ measurements.
  \vspace{-0.2cm}
  \item Lower input \ecut\ values can be more accurately constrained, while higher values require higher data quality for the same relative accuracy. Also, a lower \ecut\ leads to a higher $\tau_e\!>3$ fraction.
  \vspace{-0.15cm}
\end{itemize}
These simulation-based trends show that $\tau_e\!>3$ constraints, especially those from pre-\nustar\ literature, are a natural consequence of limited data quality.

\vspace{0.5cm}
% ---------------------------------------------------------------------------------------------
\subsection{Systematic Uncertainties} % -------------------------------------------------------
\label{sec:discussion-sysunc} % ---------------------------------------------------------------
% ---------------------------------------------------------------------------------------------

As we have shown in the preceding section using simulated data, care should be taken in interpreting any one particular measurement for an individual AGN. However, the simulations also show that at the sample level effects of random under- and overestimates average out, allowing us to more reliably constrain the median of the intrinsic \ecut\ distribution. Our method of estimating the median of the \ecut\ distribution (\S\,\ref{sec:results-ecutdist}) rests on certain assumptions; in this section we test them and discuss how our choices contribute to systematic uncertainties at the sample level. In Figure~\ref{fig:cdfs} we demonstrate the impact of a series of choices according to their effect on the CDF of the inferred \ecut\ distribution, from largest in the top panel to smallest in the bottom panel.

% ---------------------------------------------------------------------------------------------
\subsubsection{Choice of the PDFs} % ----------------------------------------------------------
\label{sec:discussion-sysunc-pdf} % -----------------------------------------------------------
% ---------------------------------------------------------------------------------------------

The largest contribution to systematic uncertainty in estimating the median \ecut\ is the choice of the range for uniform PDFs assumed for numerous lower limits. In order to demonstrate this, we define a scale parameter, $q$, so that uniform lower-limit PDFs span a range of $(q\!\times\!300)$\,keV, while for partial constraints (i.e., those lacking $E_{\rm \small cut}^{\rm \small \,u.l.}$) we assume $E_{\rm \small cut}^{\rm \small \,u.l.} = E_{\rm \small cut}^{\rm \small \,b.f.} + \left(q\!\times\!150\right)$\,keV. For our fiducial analysis presented in \S\,\ref{sec:results-ecutdist} we adopted $q\!=\!1$. In the top panel of Figure~\ref{fig:cdfs} we show the effect of assuming a narrower range ($q\!=\!0.5$) as well as a broader one ($q\!=\!2$): the median \ecut\ in those cases shifts to $230\pm10$\,keV and $390\pm40$\,keV, respectively. Although arbitrary, we argue that $q\!\approx\!1$ is a reasonable choice. A range much narrower than 300\,keV would be overly optimistic given the spread in individual constraints found from simulations presented in \S\,\ref{sec:discussion-sims}. At the same time, a range of 300\,keV is the maximum observed among individual full constraints in our sample.

The choice of PDF shape for both limits and full constraints affects the inferred \ecut\ distribution, but to a smaller extent. The effects of some alternative choices are shown in Figure~\ref{fig:cdfs}, second panel from the top. The PDFs chosen for our fiducial analysis are uniform ($U$) for lower limits and a combination of a Gaussian ($G$; below $E_{\rm \small cut}^{\rm \small \,b.f.}$) and a Lorentzian ($L$; above $E_{\rm \small cut}^{\rm \small \,b.f.}$), designated as $G|L$ in Figure~\ref{fig:cdfs}. If, instead, we employed a two-sided Gaussian ($G|G$) for full constraints, the PDF would be slightly narrower, and the sample median would shift to $250\pm20$\,keV. This can be easily understood as a consequence of the stronger high-energy tail of the Lorentzian compared to the Gaussian. Because of numerous lower limits, changing their assumed PDF produces a slightly larger effect. One could assume the same $G|L$ combination of for all constraints, in which case $E_{\rm \small cut}^{\rm \small \,b.f.}$ for lower limits could be estimated as $E_{\rm \small cut}^{\rm \small \,b.f.} = E_{\rm \small cut}^{\rm \small \,l.l.}\!+q\!\times\!150$\,keV and they would be treated as partial constraints. The median \ecut\ would then shift to $360\pm40$\,keV (assuming $q=1$ as in the fiducial case), again owing to the long high-energy tail of the Lorentzian.

% ---------------------------------------------------------------------------------------------
\subsubsection{Cuts Based on Data Quality} % --------------------------------------------------
\label{sec:discussion-sysunc-dataq} % ---------------------------------------------------------
% ---------------------------------------------------------------------------------------------

\begin{figure}[t]
\begin{center}
\vspace{0.3cm}
\includegraphics[width=1.0\columnwidth]{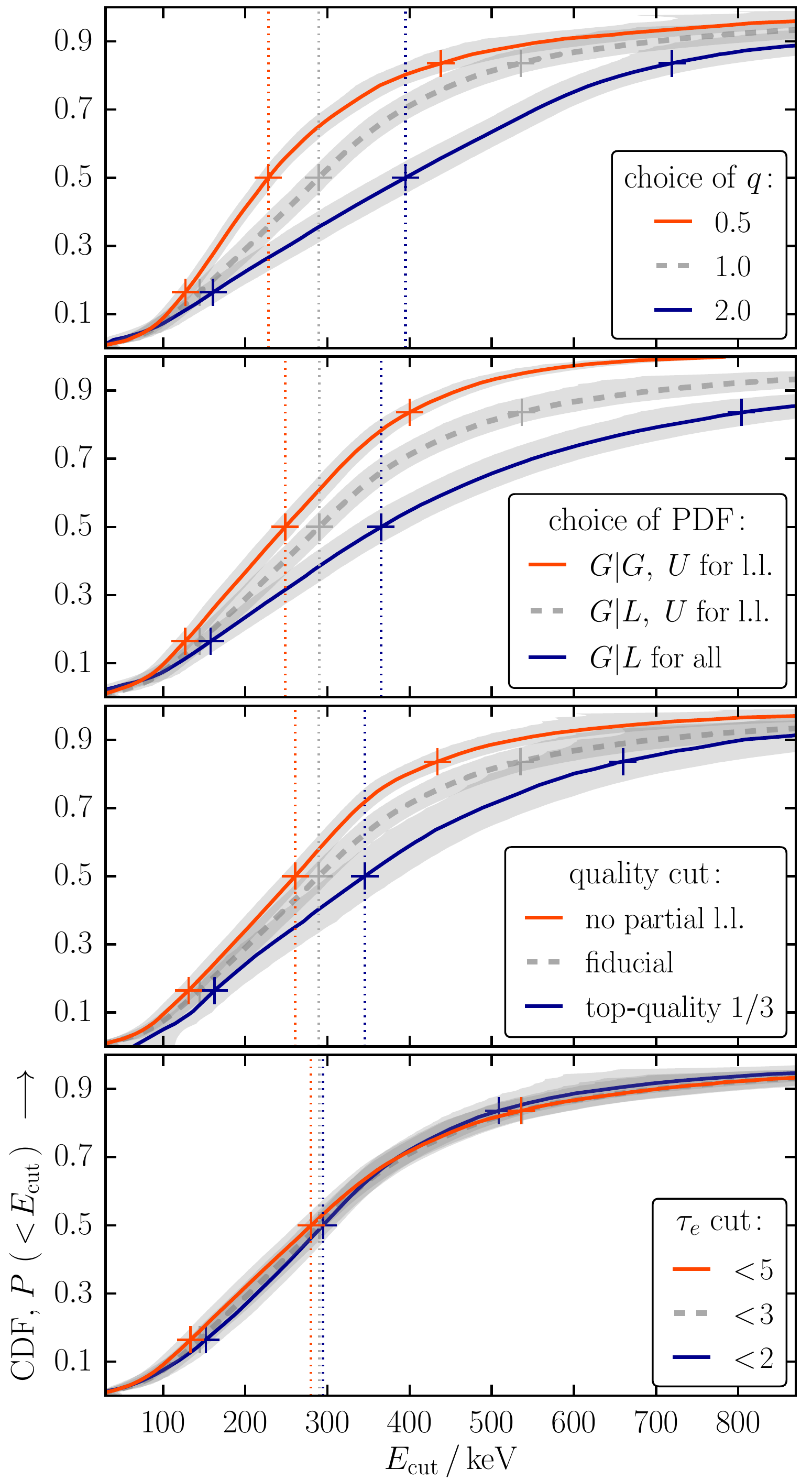}
\caption{ CDFs for \ecut\ under different assumptions. The fiducial distribution is reproduced in all panels as the thick gray dashed line. Gray areas surrounding each curve represent the 68\,\% range occupied by randomized CDFs, constructed using the procedure described in \S\,\ref{sec:results-ecutdist}. Solid lines show the median CDF in each case. Crosses intersecting each curve mark the median \ecut\ (emphasized with vertical dotted lines) and the 68\,\% interval for each CDF. From top to bottom, panels illustrate the effects of varying the PDF scale parameter ($q$), changing the assumed shape of the PDFs, applying quality-based cuts, and different cuts in $\tau_e$.
\label{fig:cdfs}}
\end{center}
\end{figure}

We consider cuts directly related to data quality by dividing our sample into thirds according to the number of degrees of freedom as a proxy of overall data quality. For the top, middle, and bottom third of the sample, we obtain \ecut\ medians at $340\pm40$\,keV, $310\pm30$\,keV, and $230\pm30$\,keV, respectively. This indicates that data quality has the tendency to lower the \ecut\ median, as expected from simulations presented in \S\,\ref{sec:discussion-sims}. Note that the fraction of $\tau_e\!>\!3$ constraints does not strongly depend on data quality (21\,\% for the top and bottom thirds, 29\,\% for the middle third), but the fraction of lower limits does: it changes from 24\,\% in the top, to 47\,\% in the middle, and 61\,\% in the bottom third. In the third panel from the top in Figure~\ref{fig:cdfs} we show the CDF constructed from the top third of the sample by data quality. The same panel also shows the effect of treating partial constraints (i.e., those lacking $E_{\rm \small cut}^{\rm \small \,u.l.}$ as lower limits instead of as full constraints. Because of their relatively small contribution, the effect on median \ecut\ is minor.

The effect of a different choice of a value for the cut in $\tau_e$ is small when applied to our sample. As shown in the bottom panel of Figure~\ref{fig:cdfs}, the median \ecut\ would shift to $300\pm20$\,keV and to $280\pm20$\,keV in the case of $\tau_e\!<\!2$ and $\tau_e\!<\!5$ cuts, respectively. With a cut in $\tau_e$ we might also be excluding AGN that might have a genuinely low \ecut, or an atypically hard intrinsic continuum ($\Gamma<1.4$). Such examples have been found in analyses of a few type~I AGN with high-quality \nustar\ data (e.g., \citealt{tortosa+2017-grs1734}, \citealt{kara+2017-ark564}, \citealt{turner+2018-1h0419}). Since a complete removal of the cut would result in lowering our sample median by only 30\,keV, we estimate that a small fraction of true low-\ecut\ AGN would lower the median by $\leq20$\,keV. We estimate that this fraction is low because our sample selection in the hard X-ray band is likely biased against AGN with soft intrinsic continua and/or low \ecut.

% ---------------------------------------------------------------------------------------------
\subsubsection{Exclusion of Individual AGN or Small Subsamples} % -----------------------------
\label{sec:discussion-sysunc-excl} % ----------------------------------------------------------
% ---------------------------------------------------------------------------------------------

This study is directed at constraining the typical \ecut\ in a sample representative of the Sy\,II population selected in the hard X-ray band. The size of our sample ensures that individual AGN do not affect the result even for the extreme examples such as Cen\,A (\ecut\,$=550_{-90}^{+140}$\,keV), NGC\,7172 (\ecut\,$>670$\,keV), or ESO\,033-G002 (\ecut\,$>460$\,keV). Excluding all three lowers the median \ecut\ by $<10$\,keV. We further consider exclusion of small subsamples of AGN that may be less representative of the Sy\,II population: those with potential contamination from a relativistic jet (broadly, {\em radio-loud} AGN), and those with a detection of at least one broad line in their optical spectra (optical type 1.9). They are marked as members of these subsamples in Table~\ref{tab:main}; further details of their membership are discussed in Paper~I. Excluding either of these subsamples, we find the median \ecut\ well within the statistical uncertainty for the full sample.

% ---------------------------------------------------------------------------------------------
\subsubsection{Alternative Estimation Method} % -----------------------------------------------
\label{sec:discussion-sysunc-alt} % -----------------------------------------------------------
% ---------------------------------------------------------------------------------------------

Finally, we also test our results using independent and more commonly used tools of ``survival analysis.'' The Kaplan-Meier (KM) estimator adapted for analysis of univariate censored data in astronomy \citep{feigelson+nelson-1985-survival1d} is a part of the \texttt{ASURV} software package \citep{feigelson+2014-asurv-code}. The particular implementation we employ here is written in Python and was first used in the study by \citet{shimizu+2016}. For our fiducial sample, the Kaplan-Meier estimate of the \ecut\ mean is 270\,keV, while the median is 230\,keV. Since only best-fit values and lower limits are taken into account in these calculations, we estimate uncertainty on the median \ecut\ from a bootstrap procedure similar to our default method (except that there is no need to assume a PDF for lower limits). Due to the asymmetry of the assumed PDF for full constraints ($G|L$; see Figure~\ref{fig:pdfs}), which gives more weight to higher energies, the median \ecut\ is then $260\pm10$\,keV. The lower bound of the 68\,\% interval is at 110\,keV with a statistical uncertainty smaller than 10\,keV.

% ---------------------------------------------------------------------------------------------
\subsubsection{Summary of Systematic Uncertainties} % -----------------------------------------
\label{sec:discussion-sysunc-summary} % -------------------------------------------------------
% ---------------------------------------------------------------------------------------------

From the analyses presented in \S\,\ref{sec:discussion-sysunc}, we conclude that the median \ecut\ for our Sy\,II sample can be constrained robustly based on our collection of individual \ecut\ constraints. We additionally verified, through simulations presented in \S\,\ref{sec:discussion-sims} and tests on data for individual AGN, that alternative choices of the fitting statistic and different data binning strategy do not systematically bias the sample-level results. Taking into account the examined systematics, we place the median \ecut\ at $290\pm50$\,keV, where $\pm20$\,keV is statistical uncertainty and $\pm30$\,keV is added as estimated systematic uncertainty. We find that the lower bound of the 68\,\% interval of the intrinsic \ecut\ distribution is relatively insensitive to various assumptions, placing it at $140\pm20$\,keV (with statistical and systematic uncertainties approximately equal). The upper end of the \ecut\ distribution is clearly less constrained by our data and therefore depends heavily on assumptions. We estimate that the upper bound of the 68\,\% interval is likely in the range of 400--600\,keV.

\vspace{1.2cm}
% ---------------------------------------------------------------------------------------------
\subsection{Comparison with the Literature} % -------------------------------------------------
\label{sec:discussion-lit} % ------------------------------------------------------------------
% ---------------------------------------------------------------------------------------------

% ---------------------------------------------------------------------------------------------
\subsubsection{Individual AGN} % --------------------------------------------------------------
\label{sec:discussion-lit-individual} % -------------------------------------------------------
% ---------------------------------------------------------------------------------------------

For the majority of targets in our sample there are no strong constraints on \ecut\ in the literature, but there is some scope for comparison in the cases of relatively bright examples. In Table~\ref{tab:litcomp} we provide an overview of published \ecut\ constraints for 36 individual AGN. The list is not exhaustive, as it only contains AGN also present in our sample, but it provides an illustrative comparison. The table contains results from the largest samples of obscured AGN studied to date, as well as studies of individual AGN or small samples based on \nustar\ data. We limit listing results from \citet{ricci+2017-bass} to targets that also have a constraint from another study; our overlap with their \swiftbat\ sample is nearly complete, but the subset given in Table~\ref{tab:litcomp} is sufficiently illustrative. We only compare to results that included a spectral model equivalent to ours, i.e., including a reprocessing component such as \pexrav.

A comparison of our results for individual AGN with those from the pre-\nustar\ literature reveals a mix of consistent and inconsistent \ecut\ constraints. Table~\ref{tab:litcomp} highlights some patterns. Of the 82 listed pre-\nustar\ constraints, 44 are lower limits that are generally consistent with our spectral fitting results. Of the remaining 38, only 12~are likely credible constraints according to our $\tau_e$-based cut defined in \S\,\ref{sec:data-degenerate}, while the majority correspond to $\tau_e\!>\!3$. Due to systematic uncertainties expected based on the simulations presented in \S\,\ref{sec:discussion-sims}, it is difficult to evaluate the significance of differences that could be ascribed to real spectral variability. In line with simulations, our results in the $\tau_e\!>\!3$ regime tend to match the previous ones for AGN with high obscuration (\lognhcmmt\,$>23.5$; e.g., Mrk\,417, NGC\,4507, NGC\,4992), strong Compton humps (\rpex\,$\gtrsim$\,1; e.g., Mrk\,1210), a combination of both (e.g., NGC\,3281), or simply limited-quality data (e.g., 2MASX\,J2330).

Comparing our spectral modeling results with previous constraints based on \nustar\ data reveals that they are not always consistent. For example, there is a stark difference for NGC\,5506 (\ecut\,$=110\pm10$\,keV) in comparison to a slightly different spectral modeling of the same \nustar\ data by \citet{matt+2015-ngc5506}, who argued that \ecut\,$>\!350$\,keV is a robust constraint (in lieu of the initial analysis, which yielded \ecut\,$=720_{-190}^{+130}$\,keV). Similarly, we find \ecut\,$=550_{-90}^{+140}$\,keV for Cen\,A, whereas \citet{fuerst+2016-cenA} concluded that \ecut\,$>\!1000$\,keV. Both cases are discussed further in the Appendix, as are the differences for AGN also studied by \citet{rani+2019} and other notably discrepant results. They clearly highlight the importance of acknowledging and quantifying systematics related to the choice of ancillary data and spectral model for any particular \ecut\ measurement even when high-quality \nustar\ data are used.

% ---------------------------------------------------------------------------------------------
\subsubsection{Sy\,II Samples} % --------------------------------------------------------------
\label{sec:discussion-lit-syII} % -------------------------------------------------------------
% ---------------------------------------------------------------------------------------------

Studying broadband spectra of local AGN with \bepposax, \citet{dadina-2007,dadina-2008} found that the average \ecut\ for their local Sy\,II sample is $380\pm40$\,keV. This is based on 25~lower limits and only 6~full constraints, but the limits were incorporated into the calculation using the KM estimator. The average \ecut\ for the full sample ($290\pm20$\,keV) is less dominated by limits and therefore likely more reliable. Systematic uncertainties related to various assumptions (e.g., cross-normalization of nonoverlapping \bepposax\ instruments, handling of lower limits) were not considered in this work, making it appear that the relative uncertainty on the average \ecut\ is the same as in our case despite the smaller sample and lower-quality data. Nevertheless, the instantaneous broadband coverage, the similarity of the spectral model used for the analysis, and the use of censored statistics make this study most directly comparable to ours out of those currently in the literature.

Studying broadband X-ray spectra of obscured AGN selected with \integral, \citet{deRosa+2012} obtained full constraints on \ecut\ for 10~targets, averaging 150\,keV. The remainder of their sample (about two-thirds) yielded lower limits below 300\,keV. Applying our method of estimating the median \ecut\ to constraints found by \citet{deRosa+2012} results in a median consistent with ours, especially if a $\tau_e\!<\!3$ cut is applied as well. We note that about a third of their constraints fall in the $\tau_e\!>\!3$ regime. This is consistent with our simulations presented in \S\,\ref{sec:discussion-sims} for true \ecut\ in the range of 250--350\,keV, with the caveat that our simulations assume lower-quality soft X-ray data. If the underlying population average were as low as 150\,keV, the fraction of $\tau_e\!>\!3$ constraints in their sample would likely be higher ($\gtrsim$50\,\%; see Figure~\ref{fig:dataqsim}).

\citet{ricci+2017-bass} derived \ecut\ constraints for many \swiftbat-selected AGN (including most of our sample) using only \swiftbat\ data in the hard X-ray band. As the majority of their constraints are lower limits, they employed a Monte Carlo scheme similar to ours in order to estimate the sample median. The main difference is the fixed upper end of the energy range for the uniform PDF assumed for lower limits instead of our scale parameter $q$ defined in \S\,\ref{sec:discussion-sysunc}. Assuming uniform PDFs extending up to 1000\,keV, for obscured AGN (\lognhcmmt\,$>\!22$) \citet{ricci+2017-bass} found the \ecut\ median at $380\pm20$\,keV. Replacing 1000\,keV with 500\,keV, which more closely corresponds to our $q=1$ assumption, lowered the median for the full sample to $240\pm10$\,keV. As a further check, they employed a KM estimator, finding the median \ecut\ for obscured AGN to be $190\pm30$\,keV. These calculations include $\tau_e\!>\!3$ constraints that our analysis would have excluded. Neglecting the $\tau_e$-based cut lowers our median by about 30\,keV, so it is reasonable to expect that incorporating such a cut would increase their estimates by at least as much. We base this estimate on the higher expected and observed fraction of $\tau_e\!>\!3$ constraints in their sample ($\simeq$35\,\%) compared to Figure~\ref{fig:dataqsim}, keeping in mind the assumed lower quality of soft X-ray data in our simulations.

Most other studies of obscured AGN samples are less directly comparable with ours. From spectral analysis of stacked \integral\ data for a local Sy\,II sample, \citet{ricci+2011} found \ecut\,$=150_{-80}^{+190}$\,keV. Subdividing their sample by obscuration and using a model that includes a \pexrav\ component, they constrained \ecut\ to $430_{-120}^{+270}$\,keV for AGN obscured by column density \lognhcmmt\,$<\!23$ and $290_{-60}^{+110}$\,keV for AGN with $23<$\,\lognhcmmt\,$<\!24$. Our simulations (\S\,\ref{sec:discussion-sims}) suggest that such a trend might arise artificially, as \ecut\ is simply more difficult to constrain at high obscuration owing to the greater parameter degeneracy. For the same subdivision by column density but using stacked \swiftbat\ data, \citet{esposito+walter-2016} successfully modeled both classes assuming \ecut\,$=250$\,keV, finding a 68\,\% lower limit around 180\,keV.

% ---------------------------------------------------------------------------------------------
\subsubsection{Sy~I and Mixed Samples} % ----------------------------------------------------
\label{sec:discussion-lit-population} % -------------------------------------------------------
% ---------------------------------------------------------------------------------------------

Studies of the \ecut\ distribution in unobscured and type~1 AGN (jointly referred to as Sy~I here) have previously been carried out using \bepposax, \integral, and \swift\ data covering the hard X-ray band. Typically focusing on samples of up to a few dozen targets, studies like \citet{perola+2002}, \citet{panessa+2008-integraltype1}, and \citet{vasudevan+2013}, for example, found a spread between 50\,keV and roughly 500\,keV. Analyses of stacked \integral\ data \citep{ricci+2011} and stacked \swiftbat\ data for the \integral-selected sample \citep{esposito+walter-2016} both found the average \ecut\ to be above 150\,keV. In a study of the average broadband X-ray AGN spectrum (constrained by luminosity functions in different X-ray bands), \citet{ballantyne-2014} indirectly found \ecut\,$=270_{-80}^{+170}$\,keV.

\citet{molina+2013} studied constraints from both \integral\ and \swiftbat\ data, estimating that the average \ecut\ over both types is in the 200--220\,keV range (with a spread of several hundred keV) when lower limits are considered and $\tau_e\!>\!3$ constraints are included. The study of \citet{ricci+2017-bass} resulted in a large collection of \ecut\ constraints for \swiftbat-selected AGN of all types. They found the median \ecut\ for the whole sample to be in the range of 240--380\,keV depending on the assumed upper end of uniform PDFs for lower limits (500--1000\,keV). Note that $q=1$ adopted for our fiducial analysis is more directly comparable to the lower end of that range, and that this median estimate would have likely been higher had $\tau_e\!>\!3$ constraints been excluded.

Not all broadband X-ray studies of obscured and unobscured AGN samples in the literature found a population-average \ecut\ consistent with our median. While some authors claimed that the average \ecut\ is likely higher than the energy range of most X-ray instruments, above 200\,keV (e.g., \citealt{dadina-2008}, \citealt{vasudevan+2013}, \citealt{ricci+2018-coronae}), others concluded that the average is likely below 150\,keV (e.g., \citealt{molina+2009}, \citealt{malizia+2014-type1cutoff}). Recent work by \citet{molina+2019}, based on \nustar\ data for the \integral-selected type~1 AGN sample previously studied by the authors, confirmed their earlier results. An independent study of a different sample by \citet{rani+2019} claimed a similarly low average \ecut. These results pointing to the typical \ecut\ below 150\,keV in Sy\,I samples are apparently inconsistent with our result for the Sy\,II population. Since they do not suffer from lower hard X-ray data quality as pre-\nustar\ studies do, this may be indicative of an emerging difference in future studies that will include a more detailed consideration of potential biases.

Within the framework of the orientation-based Unified Model for AGN (\citealt{antonucci-1993}, \citealt{urry+padovani-1995}), differences in coronal spectra can be expected if the AGN corona has a net velocity perpendicular to the accretion disk, possibly related to the formation of a relativistic jet (e.g., \citealt{malzac+2001}, \citealt{markoff+2005-jetCorona}, \citealt{liu+2014-outflowingCorona}). With the data presented herein, the estimated systematic uncertainty in the median \ecut\ of our sample, and the results currently in the literature, it is not possible to firmly establish whether the \ecut\ distributions in Sy\,I (largely unobscured) and Sy\,II (largely obscured) populations in the local universe are significantly different or not. This would require a detailed account of systematic uncertainties and selection biases, which is beyond the scope of this paper. For example, our sample selection in the hard X-ray band is likely biased against Sy\,II with soft intrinsic continua (as counterparts of the narrow-line Sy\,I population; e.g., \citealt{boller+1996-nlsy1}), but their share in the Sy\,II population is currently unknown. Furthermore, there are several limitations of our current spectral analysis that can be improved on in future work, thus enabling reliable comparisons between different subclasses. We describe some of them in the following section.

% ---------------------------------------------------------------------------------------------
\subsection{Limitations and Future Work} % ----------------------------------------------------
\label{sec:discussion-future} % ---------------------------------------------------------------
% ---------------------------------------------------------------------------------------------

The data presented here are from single-epoch, short, simultaneous \nustar\ and \swiftxrt\ observations, aided with \swiftbat\ data integrated over 70~months. An increase in \nustar\ exposure from the typical 20 ks exposure presented here to 50\,ks or 100\,ks is achievable for a subset of our sample and would yield improvement approximately summarized in Figure~\ref{fig:dataqsim}. Analyses of multiepoch data for selected unobscured (e.g., \citealt{turner+2018-1h0419}) and obscured AGN (e.g., \citealt{buisson+2018-eso103}) already provided improved constraints on spectral parameters of the coronal emission and uncovered their variability (e.g., \citealt{zoghbi+2017-mcg5}, \citealt{ursini+2018-3c382}, \citealt{zhang+2018-ecutvar}). AGN are well known to be variable, and additional multiepoch \nustar\ data would enable studies of variability in the physical parameters of the corona, the innermost parts of the accretion flow, and their scaling relations (e.g., \citealt{keek+ballantyne-2016-mrk335}).

An important consideration for characterizing coronae of obscured AGN is that the shape of the reprocessed continuum---specifically, the Compton hump---is fixed in the analysis presented here. For simplicity, we neglected possible contributions from relativistically broadened reprocessing in the inner accretion disk, focusing only on reprocessing ascribed to the larger-scale obscuring torus, although both could be contributing to the X-ray spectra of obscured AGN (e.g., \citealt{guainazzi+2010-ngc5506}, \citealt{xu+2017-iras05189}, \citealt{walton+2019-iras00521}). Despite its popularity in the literature, the \pexrav\ model we employed here does not have the correct geometry to properly represent the torus. Spectral models for reprocessing in the torus exhibit a greater range in the shape and amplitude of the Compton hump (e.g., \citealt{paltani+ricci-2017}, \citealt{buchner+2019}, \citealt{tanimoto+2019}). In future publications we will use modern spectral models from the BORUS suite \citep{balokovic+2018,balokovic+2019-rnaas} that self-consistently account for the cutoff in the coronal continuum, as well as the reprocessing features from the torus.

The simple power law with an exponential cutoff is often used in X-ray spectral analyses, and as such it is important for characterization of the average AGN spectrum (e.g., for CXB modeling). However, it is not an accurate representation of the AGN coronal spectrum (e.g., \citealt{fabian+2015}, \citealt{lubinski+2016}, \citealt{niedzwiecki+2019}). Although the phenomenological parameter \ecut\ may be approximately converted to a coronal temperature under certain assumptions \citep{middei+2019}, the proper way to characterize the physical parameters of obscured AGN coronae is to directly employ more physically motivated spectral models for coronal emission. This will enable more straightforward comparison between coronae of obscured and unobscured AGN, their physical properties, and possibly scaling relations and evolution (e.g., \citealt{kammoun+2017-oneQSOcorona}, \citealt{lanzuisi+2019}). More similar studies are expected in the future based on \nustar\ data, though higher sensitivity and higher-energy coverage of the proposed missions {\em HEX-P} \citep{madsen+2018-hexp,madsen+2019-hexpwp} or {\em FORCE} \citep{nakazawa+2018-force} will be needed in order to reach large AGN samples and detailed coronal physics \citep{kamraj+2019-wp}.

\acknowledgments

The authors appreciate helpful suggestions from the anonymous referee, which helped to improve the clarity of the paper.

M.\,B. acknowledges support from the YCAA Prize Postdoctoral Fellowship and support from the Black Hole Initiative at Harvard University, which is funded in part by the Gordon and Betty Moore Foundation (grant GBMF8273) and in part by the John Templeton Foundation. This work was funded in part by the National Aeronautics and Space Administration (NASA) under the NASA Earth and Space Science Fellowship program (grant NNX14AQ07H). A.\,C. and G.\,M. acknowledge support from the ASI/INAF grant I/037/12/0 and the Caltech Kingsley fellowship program. A.\,A. acknowledges financial support from Ministry of Education Malaysia Fundamental Research Grant Scheme grant code FRGS/1/2019/STG02/UKM/02/7. P.\,B. acknowledges financial support from the STFC and the Czech Science Foundation project No. 19-05599Y. P.\,G. acknowledges support from the STFC and a UGC/UKIERI Thematic Partnership.

We have made use of data from the \nustar\ mission, a project led by the California Institute of Technology, managed by the Jet Propulsion Laboratory, and funded by the National Aeronautics and Space Administration. We thank the \nustar\ Operations, Software and Calibration teams for support with the execution and analysis of these observations. This research has made use of the \nustar\ Data Analysis Software (NuSTARDAS) jointly developed by the Space Science Data Center (SSDC; ASI, Italy) and the California Institute of Technology (USA). Part of this work is based on archival data, software or online services provided by the SSDC. This research has made use of the High Energy Astrophysics Science Archive Research Center Online Service, provided by the NASA/Goddard Space Flight Center and NASA's Astrophysics Data System.

\facilities{\nustar, \swift}

\software{\texttt{Astropy} \citep{astropy-2013,astropy-2018}, \texttt{ASURV} \citep{feigelson+2014-asurv-code}, \texttt{Matplotlib} \citep{hunter+2007}, \xspec\ \citep{arnaud-1996}}

% FULL TABLE WITH CONSTRAINTS:

\input{tab01.tex}

% COMPARISON WITH LITERATURE:

\input{tab02.tex}

\appendix

%%%%%%%%%%%%%%%%%%%%%%%%%%%%%%%%%%%%%%%%%%%%%%%%%%%%%%%%%%%%%%%%%%%%%%%%%%%%%%%%%%%%%%%%%%%%%%%
\section{Notes on Particular Targets} %%%%%%%%%%%%%%%%%%%%%%%%%%%%%%%%%%%%%%%%%%%%%%%%%%%%%%%%%
\label{sec:appendix} %%%%%%%%%%%%%%%%%%%%%%%%%%%%%%%%%%%%%%%%%%%%%%%%%%%%%%%%%%%%%%%%%%%%%%%%%%
%%%%%%%%%%%%%%%%%%%%%%%%%%%%%%%%%%%%%%%%%%%%%%%%%%%%%%%%%%%%%%%%%%%%%%%%%%%%%%%%%%%%%%%%%%%%%%%

{\bf NGC\,2110} -- While the high-\ecut\ constraint from \citet{ricci+2017-bass} is not strongly inconsistent with our analysis, we confirm that our lower \nustar-based constraint is robust despite achieving no improvement in terms of $\chi^2$ reduction with respect to modeling presented in Paper~I. Using \integral\ data and two epochs of \nustar\ data in the hard X-ray band, \citet{ursini+2019-ngc2110+4388} affirm our result, surpassing an earlier lower limit ($>\!210$\,keV) found by \citet{marinucci+2015-ngc2110}.

{\bf NGC\,2992} -- \nustar\ observation of this target was taken at a time when its hard X-ray flux was more than a factor of~5 higher than the long-term average probed by \swiftbat. High flux was also identified in the multiepoch spectral analysis by \citet{marinucci+2018-ngc2992}, which yielded \ecut\,$>350$\,keV. Excluding the highly offset \swiftbat\ data results in an even higher lower limit, we find \ecut$>\!490$\,keV. Lowering \ecut\ manually, we find that statistically acceptable models (\pnull\,$>\!5$\,\%) can be found down to 80\,keV (formally consistent with the result from \citealt{rani+2019}), albeit with clear excess in residuals at the high-energy end of the \nustar\ bandpass.

{\bf MCG\,--05-23-016} --- In addition to the \nustar-based study of \citet{balokovic+2015-mcg5} listed in Table~\ref{tab:litcomp}, \nustar\ data for this bright target were also analyzed by \citet{zoghbi+2017-mcg5} and \citet{tortosa+2018-sample}. While each of those analyses implies a different \ecut\ owing to details of data selection and spectral models used for fitting, there is agreement that in this case \ecut\ is likely $\lesssim150$\,keV, i.e., low enough that \nustar\ data constrain it well.

{\bf NGC\,4388} --- For this target our constraint agrees very well with a more detailed study by \citet{ursini+2019-ngc2110+4388}, including \nustar\ and \integral\ hard X-ray data. We also confirm \ecut\,$\approx\,200$\,keV from a dedicated multiepoch study with a different spectral model (M. Balokovi\'{c} et al., in prep.). The data used in our analysis do not strongly reject \ecut\ above lower limits found by \citet{deRosa+2012} and \citet{vasudevan+2013} as an acceptable model with \pnull\,$>\!5$\,\% can be found as long as \ecut\,$>160$\,keV.

{\bf Cen\,A} --- The discrepancy between the \swiftbat-based constraint from \citet{ricci+2017-bass}, which is in the $\tau_e\!>\!3$ regime ($70\pm10$\,keV), and nearly all other studies implying a very high \ecut\ (including this paper and \citealt{fuerst+2016-cenA}, based on high-quality \nustar\ data), is probably the best example of the spurious occurrence of degenerate $\tau_e\!>\!3$ constraints. The cause for this may be the curvature of the \swiftbat\ spectrum, which does not match the very flat \nustar\ spectrum. For this reason we excluded \swiftbat\ data from our fitting already in Paper~I as it was impossible to find an acceptable model otherwise. The extremely high \ecut\ may be indicative of a contribution from the relativistic jet in this AGN, so we test its exclusion from the sample in \S\,\ref{sec:discussion-sysunc-excl}, finding negligible impact on the population median.

{\bf NGC\,5506} --- Initial analysis of \nustar\ data for this target by \citet{matt+2015-ngc5506}, as well as a reanalysis by \citet{tortosa+2018-sample} using a slightly different spectral model (replacing the \pexrav\ component with the \texttt{xillver} model from \citealt{garcia+2013}, which self-consistently includes fluorescent line emission), yielded a constraint on \ecut\ far above the \nustar\ band, at $\simeq700$\,keV. A more robust lower limit, found by considering alternative models, was found to be $>\!350$\,keV. However, \ecut\ in that range is inconsistent with the highly curved \swiftbat\ spectrum. Excluding it from our fitting, we find \ecut\,$\,=210_{-30}^{+120}$\,keV. Also excluding \swiftbat\ data and using spectral models self-consistently accounting for coronal emission and reprocessing, \citet{balokovic+2019-rnaas} found $kT_e$ in the 70--110\,keV range (depending on the chosen model), which corresponds to \ecut$\approx140\!-\!330$\,keV assuming the simple conversion from Equation~\ref{eq:kt}.

{\bf NGC\,6300} --- The only apparent discrepancy in constraints for this target is with respect to joint modeling of \integral\ and \swiftbat\ spectra by \citet{molina+2013}. The high lower limit ($>\!520$\,keV) is based on their modeling with a free \rpex\ parameter and is likely spurious. From fits with \rpex\ fixed at 0, 1, and 2, they derive \ecut\ in the range of 40--90\,keV, all of which correspond to $\tau_e\!>\!3$.

{\bf NGC\,7172} --- Removing the \swiftbat\ data from the fit does not change the high lower limit on \ecut, and neither does further exclusion of \swiftxrt\ data. We find a more robust lower limit, such that no models meet the \pnull\,$>5\,\%$ criterion below it, at \ecut\,$>250$\,keV. We are unable to match the data with the spectral parameters listed by \citet{rani+2019}, who constrained \ecut\ to $70\pm10$\,keV based on \nustar\ data alone. A similarly low and tight \ecut\ constraint was found by \citet{deRosa+2012} from nonsimultaneous and nonoverlapping \xmmnewton\ and \integral\ data. In both cases these constraints correspond to $\tau_e\!>\!3$.

\bibliographystyle{aasjournal} % .bst file
\bibliography{bib} % .bib file

\end{document}

%% file: tab01.tex
\startlongtable
\begin{deluxetable*}{lcccccc}
\tablecaption{Constraints on parameters of the hard X-ray continuum model. \label{tab:main}}
\tabletypesize{\small}
\tablehead{
  \colhead{Target Name} &
  \colhead{$\chi^2/\nu$} &
  \colhead{\pnull\,/\,\%} &
  \colhead{$\Gamma$} &
  \colhead{\ecut$\,/\,$keV} &
  \colhead{$\tau_e$\,\tablenotemark{a}} &
  \colhead{Notes\,\tablenotemark{b}}
}
\startdata
NGC\,262 \dotfill & 503.0\,/\,464 & 10 & $1.66\pm0.04$ & $170_{-30}^{+40}$ & 2.1 & RL,\,1.9 \\
ESO\,195-IG021 \dotfill & 224.7\,/\,249 & 86 & $1.88\pm0.09$ & $>\!230$ & 1.0 & \nodata \\
NGC\,454\,E \dotfill & 56.2\,/\,79 & 98 & $1.6_{-0.5}^{+0.3}$ & $>\!50$ & {\bf 5.4} & \nodata \\
NGC\,513 \dotfill & 126.5\,/\,140 & 79 & $1.77_{-0.15}^{+0.03}$ & $>\!230$ & 1.2 & \nodata \\
NGC\,612 \dotfill & 102.1\,/\,126 & 94 & $1.47_{-0.09}^{+0.25}$ & $170_{-50}^{+u}$ & 2.8 & RL \\
2MASX\,J0140 \dotfill & 195.1\,/\,235 & 97 & $1.59\pm0.09$ & $70_{-20}^{+40}$ & {\bf 4.4} & \nodata \\
MCG\,$-$01-05-047 \dotfill & 98.9\,/\,114 & 84 & $1.7\pm0.2$ & $200_{-100}^{+u}$ & 1.8 & \nodata \\
NGC\,788 \dotfill & 121.1\,/\,149 & 96 & $1.7\pm0.2$ & $200_{-100}^{+u}$ & 1.4 & \nodata \\
ESO\,416-G002 \dotfill & 147.3\,/\,156 & 68 & $1.68\pm0.03$ & $>\!480$ & 0.7 & 1.9 \\
NGC\,1052 \dotfill & 179.9\,/\,211 & 94 & $1.36\pm0.09$ & $80_{-20}^{+40}$ & {\bf 5.4} & RL,\,1.9 \\
2MFGC\,2280 \dotfill & 24.9\,/\,37 & 94 & $1.4\pm0.5$ & $130_{-80}^{+u}$ & {\bf 3.8} & \nodata \\
NGC\,1194 \dotfill & 163.8\,/\,175 & 72 & $1.5\pm0.2$ & $140_{-60}^{+100}$ & {\bf 3.3} & \nodata \\
NGC\,1229 \dotfill & 112.8\,/\,100 & 18 & $1.6_{-0.4}^{+0.1}$ & $>\!82.4$ & {\bf 3.7} & C \\
MCG\,$+$00-09-042 \dotfill & 249.1\,/\,248 & 47 & $2.07_{-0.10}^{+0.03}$ & $>\!190$ & 0.9 & \nodata \\
NGC\,1365 \dotfill & 436.8\,/\,428 & 37 & $1.90\pm0.08$ & $290_{-100}^{+200}$ & 0.8 & \nodata \\
2MASX\,J0356 \dotfill & 177.9\,/\,155 & 10 & $1.69_{-0.09}^{+0.04}$ & $>\!240$ & 1.4 & \nodata \\
3C\,105 \dotfill & 90.1\,/\,92 & 54 & $1.8\pm0.3$ & $>\!70$ & {\bf 3.5} & C,\,RL \\
2MASX\,J0423 \dotfill & 115.3\,/\,119 & 58 & $1.4_{-0.4}^{+0.2}$ & $70_{-30}^{+40}$ & {\bf 5.5} & \nodata \\
MCG\,$+$03-13-001 \dotfill & 93.3\,/\,88 & 33 & $1.9_{-0.4}^{+0.1}$ & $>\!60$ & {\bf 3.4} & C,\,1.9 \\
CGCG\,420-015 \dotfill & 267.9\,/\,244 & 14 & $1.8\pm0.2$ & $190_{-90}^{+u}$ & 1.4 & C \\
ESO\,033-G002 \dotfill & 379.5\,/\,428 & 96 & $2.20\pm0.07$ & $>\!460$ & 0.4 & \nodata \\
LEDA\,178130 \dotfill & 374.5\,/\,384 & 63 & $1.68_{-0.06}^{+0.08}$ & $350_{-150}^{+u}$ & 0.9 & \nodata \\
2MASX\,J0508 \dotfill & 352.7\,/\,321 & 11 & $1.71\pm0.07$ & $160_{-60}^{+200}$ & 2.1 & 1.9 \\
ESO\,553-G043 \dotfill & 314.1\,/\,327 & 69 & $1.71\pm0.06$ & $>\!190$ & 1.8 & \nodata \\
NGC\,2110 \dotfill & 687.2\,/\,661 & 23 & $1.64\pm0.02$ & $300_{-30}^{+50}$ & 1.1 & A \\
ESO\,005-G004 \dotfill & 60.9\,/\,56 & 30 & $1.83_{-0.05}^{+0.18}$ & $>\!140$ & 2.1 & B \\
Mrk\,3 \dotfill & 539.3\,/\,541 & 51 & $1.52\pm0.08$ & $150_{-30}^{+60}$ & 2.8 & 1.9 \\
ESO\,121-IG028 \dotfill & 156.3\,/\,186 & 94 & $1.87_{-0.17}^{+0.09}$ & $>\!150$ & 1.7 & \nodata \\
LEDA\,549777 \dotfill & 98.5\,/\,104 & 64 & $1.4\pm0.1$ & $>\!90$ & {\bf 4.4} & \nodata \\
LEDA\,511628 \dotfill & 185.5\,/\,208 & 87 & $1.6\pm0.1$ & $90_{-30}^{+80}$ & {\bf 3.4} & \nodata \\
MCG\,$+$06-16-028 \dotfill & 101.8\,/\,91 & 20 & $1.8_{-0.3}^{+0.1}$ & $>\!110$ & 2.4 & 1.9 \\
IRAS\,07378$-$3136 \dotfill & 226.2\,/\,229 & 54 & $1.3\pm0.2$ & $60_{-20}^{+40}$ & {\bf 6.8} & \nodata \\
UGC\,3995\,A \dotfill & 167.9\,/\,188 & 85 & $1.5\pm0.2$ & $100_{-40}^{+110}$ & {\bf 3.6} & \nodata \\
Mrk\,1210 \dotfill & 284.3\,/\,286 & 52 & $1.6\pm0.1$ & $90_{-20}^{+40}$ & {\bf 3.6} & 1.9 \\
MCG\,$-$01-22-006 \dotfill & 350.3\,/\,367 & 73 & $1.4\pm0.1$ & $110_{-30}^{+60}$ & {\bf 4.7} & \nodata \\
CGCG\,150-014 \dotfill & 105.3\,/\,104 & 44 & $1.78_{-0.33}^{+0.05}$ & $>\!110$ & 2.5 & RL \\
MCG\,$+$11-11-032 \dotfill & 60.9\,/\,64 & 59 & $1.97_{-0.04}^{+0.16}$ & $>\!140$ & 1.7 & \nodata \\
2MASX\,J0903 \dotfill & 60.5\,/\,57 & 35 & $1.9\pm0.2$ & $>\!270$ & 0.9 & C \\
2MASX\,J0911 \dotfill & 116.8\,/\,133 & 84 & $1.50\pm0.09$ & $70_{-20}^{+60}$ & {\bf 4.9} & \nodata \\
IC\,2461 \dotfill & 243.3\,/\,247 & 56 & $1.8\pm0.1$ & $200_{-90}^{+u}$ & 1.4 & \nodata \\
MCG\,$-$01-24-012 \dotfill & 296.8\,/\,327 & 88 & $1.93\pm0.09$ & $110_{-30}^{+50}$ & 2.2 & \nodata \\
2MASX\,J0923 \dotfill & 82.0\,/\,85 & 57 & $1.1\pm0.6$ & $40_{-20}^{+90}$ & {\bf $>$\,7} & \nodata \\
NGC\,2992 \dotfill & 653.7\,/\,719 & 96 & $1.74\pm0.02$ & $>\!380$ & 0.8 & A,\,1.9 \\
MCG\,$-$05-23-016 \dotfill & 811.4\,/\,780 & 21 & $1.93\pm0.02$ & $150\pm10$ & 1.6 & A,\,1.9 \\
NGC\,3079 \dotfill & 91.7\,/\,102 & 76 & $1.1\pm0.4$ & $40_{-10}^{+20}$ & {\bf $>$\,7} & \nodata \\
ESO\,263-G013 \dotfill & 136.3\,/\,139 & 55 & $1.7\pm0.2$ & $>\!120$ & 2.7 & \nodata \\
NGC\,3281 \dotfill & 176.1\,/\,199 & 88 & $1.25_{-0.11}^{+0.09}$ & $70\pm10$ & {\bf $>$\,7} & \nodata \\
MCG\,$+$12-10-067 \dotfill & 130.3\,/\,138 & 67 & $2.0_{-0.3}^{+0.1}$ & $>\!108.7$ & 2.1 & \nodata \\
MCG\,$+$06-24-008 \dotfill & 139.1\,/\,146 & 64 & $1.60_{-0.08}^{+0.05}$ & $>\!170$ & 2.3 & \nodata \\
UGC\,5881 \dotfill & 95.9\,/\,94 & 43 & $1.3\pm0.3$ & $80_{-30}^{+120}$ & {\bf 5.8} & \nodata \\
NGC\,3393 \dotfill & 57.8\,/\,77 & 95 & $1.8_{-0.3}^{+0.2}$ & $160_{-100}^{+u}$ & 2.0 & \nodata \\
Mrk\,417 \dotfill & 219.0\,/\,218 & 47 & $1.51\pm0.02$ & $130_{-40}^{+120}$ & {\bf 3.2} & \nodata \\
2MASX\,J1136 \dotfill & 270.3\,/\,259 & 30 & $2.02\pm0.03$ & $>\!350$ & 0.6 & \nodata \\
NGC\,3822 \dotfill & 96.8\,/\,114 & 88 & $1.7\pm0.1$ & $>\!70$ & {\bf 3.8} & \nodata \\
B2\,1204$+$34 \dotfill & 177.5\,/\,207 & 93 & $1.70\pm0.05$ & $>\!280$ & 1.1 & RL \\
IRAS\,12074$-$4619 \dotfill & 169.0\,/\,180 & 71 & $1.85_{-0.06}^{+0.03}$ & $>\!320$ & 0.8 & 1.9 \\
WAS\,49 \dotfill & 99.2\,/\,99 & 48 & $1.4\pm0.3$ & $60_{-20}^{+60}$ & {\bf 6.2} & \nodata \\
NGC\,4258 \dotfill & 240.6\,/\,257 & 76 & $1.82_{-0.12}^{+0.05}$ & $>\!180$ & 1.5 & 1.9 \\
NGC\,4388 \dotfill & 346.5\,/\,311 & 8 & $1.64_{-0.06}^{+0.08}$ & $210_{-40}^{+120}$ & 1.8 & A,\,X \\
NGC\,4395 \dotfill & 276.3\,/\,305 & 88 & $1.54\pm0.08$ & $120_{-30}^{+50}$ & {\bf 3.3} & 1.9 \\
NGC\,4507 \dotfill & 662.7\,/\,715 & 92 & $1.41_{-0.11}^{+0.09}$ & $80_{-10}^{+20}$ & {\bf 5.2} & 1.9 \\
LEDA\,170194 \dotfill & 130.3\,/\,140 & 71 & $1.79_{-0.12}^{+0.04}$ & $>\!230$ & 1.1 & \nodata \\
NGC\,4941 \dotfill & 58.6\,/\,57 & 42 & $1.5_{-0.5}^{+0.3}$ & $110_{-60}^{+u}$ & {\bf 3.6} & \nodata \\
NGC\,4939 \dotfill & 132.4\,/\,135 & 55 & $1.8\pm0.2$ & $>\!140$ & 2.1 & \nodata \\
NGC\,4945 \dotfill & 499.8\,/\,473 & 19 & $1.91\pm0.03$ & $>\!240$ & 0.9 & B \\
NGC\,4992 \dotfill & 283.1\,/\,248 & 6 & $1.2\pm0.4$ & $80_{-30}^{+90}$ & {\bf $>$\,7} & \nodata \\
Mrk\,248 \dotfill & 141.3\,/\,166 & 92 & $1.6\pm0.1$ & $50_{-10}^{+20}$ & {\bf 5.5} & \nodata \\
Cen\,A \dotfill & 727.2\,/\,698 & 22 & $1.765\pm0.007$ & $550_{-90}^{+140}$ & 0.5 & A,\,B,\,RL \\
ESO\,509-IG066 \dotfill & 199.2\,/\,224 & 88 & $1.5\pm0.1$ & $70_{-20}^{+50}$ & {\bf 4.4} & 1.9 \\
NGC\,5252 \dotfill & 299.7\,/\,292 & 36 & $1.67\pm0.04$ & $330_{-100}^{+150}$ & 0.9 & X \\
2MASX\,J1410 \dotfill & 145.7\,/\,123 & 8 & $1.8\pm0.2$ & $>\!80$ & {\bf 3.3} & \nodata \\
NGC\,5506 \dotfill & 197.9\,/\,181 & 18 & $1.79\pm0.02$ & $110\pm10$ & 2.7 & A,\,1.9 \\
NGC\,5643 \dotfill & 125.8\,/\,115 & 23 & $1.9\pm0.2$ & $>\!130$ & 2.0 & C \\
NGC\,5674 \dotfill & 244.0\,/\,258 & 73 & $1.86\pm0.09$ & $210_{-110}^{+u}$ & 1.1 & \nodata \\
Mrk\,477 \dotfill & 183.4\,/\,189 & 60 & $1.6_{-0.1}^{+0.2}$ & $140_{-60}^{+200}$ & 2.6 & 1.9 \\
NGC\,5728 \dotfill & 271.4\,/\,270 & 46 & $1.3_{-0.2}^{+0.1}$ & $80_{-20}^{+30}$ & {\bf 6.9} & 1.9 \\
IC\,4518A \dotfill & 125.9\,/\,130 & 59 & $1.8_{-0.1}^{+0.2}$ & $120_{-50}^{+150}$ & 2.3 & \nodata \\
2MASX\,J1506 \dotfill & 78.2\,/\,94 & 88 & $1.71_{-0.09}^{+0.06}$ & $>\!140$ & 2.4 & \nodata \\
NGC\,5899 \dotfill & 227.3\,/\,239 & 70 & $1.96_{-0.05}^{+0.08}$ & $>\!340$ & 0.6 & \nodata \\
MCG\,$+$11-19-006 \dotfill & 85.2\,/\,72 & 14 & $1.5_{-0.4}^{+0.2}$ & $>\!60$ & {\bf 5.5} & 1.9 \\
MCG\,$-$01-40-001 \dotfill & 159.8\,/\,210 & 99 & $1.8\pm0.1$ & $260_{-130}^{+u}$ & 1.0 & 1.9 \\
NGC\,5995 \dotfill & 394.8\,/\,393 & 46 & $2.02_{-0.01}^{+0.05}$ & $>\!340$ & 0.6 & 1.9 \\
MCG\,$+$14-08-004 \dotfill & 92.0\,/\,99 & 68 & $1.7\pm0.2$ & $>\!120$ & 2.6 & \nodata \\
Mrk\,1498 \dotfill & 325.3\,/\,307 & 23 & $1.35_{-0.05}^{+0.10}$ & $60\pm10$ & {\bf 6.8} & RL \\
IRAS\,16288$+$3929 \dotfill & 54.2\,/\,66 & 85 & $1.7\pm0.3$ & $130_{-70}^{+u}$ & 2.7 & \nodata \\
ESO\,137-G034 \dotfill & 81.4\,/\,84 & 56 & $1.81_{-0.25}^{+0.05}$ & $>\!160$ & 1.8 & \nodata \\
LEDA\,214543 \dotfill & 294.4\,/\,340 & 96 & $1.83\pm0.09$ & $>\!170$ & 1.6 & \nodata \\
NGC\,6240 \dotfill & 309.9\,/\,319 & 63 & $1.4\pm0.2$ & $90_{-30}^{+70}$ & {\bf 4.6} & 1.9 \\
NGC\,6300 \dotfill & 338.2\,/\,303 & 8 & $1.85\pm0.06$ & $210_{-50}^{+100}$ & 1.1 & A \\
MCG\,$+$07-37-031 \dotfill & 203.9\,/\,192 & 26 & $1.66\pm0.09$ & $200_{-90}^{+u}$ & 1.9 & \nodata \\
2MASX\,J1824 \dotfill & 152.2\,/\,162 & 70 & $1.8_{-0.2}^{+0.1}$ & $>\!110$ & 2.5 & 1.9 \\
IC\,4709 \dotfill & 212.5\,/\,218 & 59 & $1.8\pm0.2$ & $140_{-60}^{+200}$ & 2.1 & \nodata \\
LEDA\,3097193 \dotfill & 376.4\,/\,392 & 70 & $1.78\pm0.07$ & $130_{-40}^{+110}$ & 2.3 & \nodata \\
ESO\,103-G035 \dotfill & 327.9\,/\,349 & 78 & $1.78\pm0.05$ & $100_{-10}^{+20}$ & 2.7 & 1.9 \\
ESO\,231-G026 \dotfill & 186.1\,/\,234 & 99 & $1.67_{-0.11}^{+0.03}$ & $>\!250$ & 1.3 & \nodata \\
2MASX\,J1926 \dotfill & 76.5\,/\,90 & 84 & $1.9\pm0.2$ & $150_{-80}^{+u}$ & 1.8 & \nodata \\
2MASX\,J1947 \dotfill & 245.0\,/\,252 & 61 & $1.8\pm0.1$ & $120_{-40}^{+110}$ & 2.5 & \nodata \\
3C\,403 \dotfill & 168.0\,/\,180 & 73 & $1.5\pm0.2$ & $>\!110$ & {\bf 3.6} & RL \\
Cyg\,A \dotfill & 629.3\,/\,633 & 53 & $1.58\pm0.08$ & $130_{-30}^{+70}$ & 2.9 & B,\,RL \\
2MASX\,J2006 \dotfill & 102.0\,/\,113 & 76 & $1.9\pm0.2$ & $>\!80$ & 2.6 & \nodata \\
2MASX\,J2018 \dotfill & 240.2\,/\,253 & 71 & $1.6\pm0.2$ & $170_{-70}^{+u}$ & 2.4 & \nodata \\
2MASX\,J2021 \dotfill & 134.1\,/\,133 & 46 & $1.7\pm0.2$ & $>\!90$ & {\bf 3.1} & \nodata \\
NGC\,6921 \dotfill & 42.2\,/\,68 & 99 & $1.7\pm0.2$ & $190_{-90}^{+u}$ & 1.7 & C \\
MCG\,$+$04-48-002 \dotfill & 70.2\,/\,74 & 60 & $1.7\pm0.2$ & $>\!150$ & 2.3 & C \\
IC\,5063 \dotfill & 291.5\,/\,287 & 42 & $1.7_{-0.1}^{+0.2}$ & $220_{-90}^{+u}$ & 1.3  & RL \\
NGC\,7130 \dotfill & 82.1\,/\,72 & 20 & $1.91_{-0.30}^{+0.07}$ & $>\!100$ & 2.4 & 1.9 \\
NGC\,7172 \dotfill & 774.5\,/\,769 & 44 & $1.88\pm0.02$ & $>\!670$ & 0.4 & A \\
MCG\,$+$06-49-019 \dotfill & 59.3\,/\,60 & 50 & $1.57_{-0.08}^{+0.06}$ & $>\!200.0$ & 2.2 & 1.9 \\
NGC\,7319 \dotfill & 95.7\,/\,93 & 40 & $1.71_{-0.18}^{+0.04}$ & $>\!220$ & 1.4 & \nodata \\
3C\,452 \dotfill & 357.1\,/\,418 & 99 & $1.3\pm0.1$ & $70_{-20}^{+40}$ & {\bf 6.2} & RL,\,1.9 \\
NGC\,7582 \dotfill & 294.4\,/\,298 & 55 & $1.7\pm0.2$ & $200_{-80}^{+190}$ & 1.8 & \nodata \\
2MASX\,J2330 \dotfill & 71.1\,/\,78 & 70 & $1.7\pm0.3$ & $>\!70$ & {\bf 3.7} & \nodata \\
PKS\,2331-240 \dotfill & 304.6\,/\,278 & 13 & $1.80\pm0.05$ & $220_{-110}^{+u}$ & 1.2 & RL \\
PKS\,2356-61 \dotfill & 127.6\,/\,157 & 96 & $1.7\pm0.2$ & $160_{-80}^{+u}$ & 2.1 & RL \\
\enddata
\tablenotetext{a}{Optical depth approximately calculated using Equation~\ref{eq:tauline} and best-fit spectral parameters $\Gamma$ and \ecut\ (or lower limit if best-fit value for the latter is undefined). Values greater than three are highlighted in bold font.}
\tablenotetext{b}{Notes: A\,$=$\,details discussed in Appendix~\ref{sec:appendix}; B\,$=$\,\swiftbat\ data excluded from spectral fitting; C\,$=$\,\nustar\ or \swiftxrt\ data co-added; X\,$=$\,\swiftxrt\ data not included; RL\,$=$\,radio-loud AGN (see \S\,\ref{sec:discussion-sysunc-excl}); 1.9\,$=$\,optical type 1.9 according to \citet{koss+2017-bass}.}
\end{deluxetable*}

%% file: tab02.tex
\begin{deluxetable*}{lCCCCCCC}
\tablecaption{ Comparison of constraints on \ecut\ for individual AGN from the literature.
\label{tab:litcomp}}
\tabletypesize{\small}
\tablehead{
  \colhead{\multirow{2}{*}{Target Name}} &
  \colhead{} & \colhead{} & \colhead{} & \colhead{\ecut$\,/\,$keV} & \colhead{} & \colhead{} & \colhead{} \\
  \cline{2-8}
  \colhead{} & \colhead{This Work} & \colhead{D07} & \colhead{dR12} & \colhead{M13} & \colhead{V13} & \colhead{R17} & \colhead{\nustar\ [Ref.]}
}
\startdata
NGC\,262 \dotfill & 170_{-20}^{+40} & \nodata & \nodata & \nodata & \nodata & \left(\,110_{-30}^{+60}\,\right) & \left(\,80_{-20}^{+40}\,\right)~\mbox{[R19]} \\
NGC\,788 \dotfill & 220_{-100}^{+u} & \nodata & \left(\,60_{-20}^{+40}\,\right) & \nodata & \nodata & >\!70 & \nodata \\
NGC\,1365 \dotfill & 290_{-100}^{+200} & >\!110 & \nodata & \nodata & \nodata & 140_{-60}^{+100} & \nodata \\
{\bf NGC\,2110} \dotfill & 300_{-30}^{+50} & >\!70 & \nodata & \nodata & \nodata & 450\pm60 & 320_{-60}^{+100}~\mbox{[U19]} \\
Mrk\,3 \dotfill & 150_{-30}^{+70} & >\!190 & >\!200 & \nodata & \nodata & >\!450 &\nodata \\
Mrk\,1210 \dotfill & \left(\,90_{-20}^{+30}\,\right) & >\!110 & \nodata & \nodata & \nodata & >\!120 & \nodata \\
MCG\,--01-24-012 \dotfill & 110_{-30}^{+60} & >\!420 & \nodata & \nodata & \nodata & \left(\,80_{-30}^{+90}\,\right) & \nodata \\
{\bf NGC\,2992} \dotfill & >\!380 & 150_{-50}^{+170} & \nodata & \nodata & \nodata & \left(\,60_{-20}^{+40}\,\right) & 150_{-70}^{+130}~\mbox{[R19]} \\
{\bf MCG\,--05-23-016} \dotfill & 150_{-20}^{+10} & 190_{-60}^{+10} & >\!170 & \left(\,70_{-30}^{+150}\,\right) & \nodata & 100\pm10 & 120\pm10~\mbox{[B15]} \\
ESO\,263-G013 \dotfill & >\!120 & \nodata & >\!150 & \nodata & \nodata & >\!60 & \nodata \\
Mrk\,417 \dotfill & \left(\,130_{-50}^{+110}\,\right) & \nodata & \nodata & \nodata & \left(\,40_{-20}^{+10}\,\right) & >\!120 & \nodata \\
NGC\,3281 \dotfill & \left(\,70_{-10}^{+20}\,\right) & \left(\,70\pm30\,\right) & >\!60 & \left(\,40_{-20}^{+120}\,\right) & \nodata & \left(\,60_{-10}^{+20}\,\right) & \nodata \\
NGC\,4258 \dotfill & >\!180 & \nodata & \nodata & \nodata & >\!280 & >\!310 & \nodata \\
{\bf NGC\,4388} \dotfill & 210_{-40}^{+120} & >\!460 & >\!180 & 200_{-150}^{+200} & \gg\!200 & >\!100 & 200_{-40}^{+80}~\mbox{[U19]} \\
NGC\,4395 \dotfill & \left(\,120_{-30}^{+50}\,\right) & \nodata & \nodata & \nodata & \left(\,50_{-10}^{+30}\,\right) & >\!260 & \nodata \\
NGC\,4507 \dotfill & \left(\,80_{-20}^{+10}\,\right) & >\!80 & 130_{-50}^{+150} & >\!60 & \nodata & 130_{-40}^{+90} &  \nodata \\
LEDA\,170194 \dotfill & >\!230 & \nodata & >\!210 & \nodata & \nodata & >\!200 & \nodata \\
NGC\,4945 \dotfill & >\!240 & 120_{-30}^{+40} & >\!80 & 100_{-60}^{+260} & \nodata & >\!40 & 190_{-40}^{+200}~\mbox{[P14]} \\
NGC\,4992 \dotfill & \left(\,80_{-30}^{+100}\,\right) & \nodata & \left(\,110_{-20}^{+190}\,\right) & \nodata & \nodata & \left(\,70_{-30}^{+140}\,\right) & \nodata \\
{\bf Cen\,A} \dotfill & 550_{-90}^{+140} & >\!250 & >\!400 & \nodata & \nodata & \left(\,70\pm10\,\right) & >\!1000~\mbox{[F16]} \\
NGC\,5252 \dotfill & 330_{-100}^{+150} & \nodata & >\!50 & \nodata & 110_{-20}^{+60} & \left(\,80\pm10\,\right) & \nodata \\
{\bf NGC\,5506} \dotfill & 110\pm10 & >\!180 & \nodata & >\!90 & 170_{-30}^{+110} & 130\pm10 & >\!350~\mbox{[M15]} \\
NGC\,5899 \dotfill & >\!340 & \nodata & \nodata & \nodata & >\!210 & >\!40 & \nodata \\
IC\,4518A \dotfill & 120_{-50}^{+160} & \nodata & \left(\,70_{-30}^{+60}\,\right) & \left(\,20_{-10}^{+60}\,\right) & \nodata & \left(\,70_{-30}^{+330}\,\right) & \nodata \\
Mrk\,1498 \dotfill & \left(\,60\pm10\,\right) & \nodata & \nodata & \nodata & \nodata & \left(\,70_{-20}^{+90}\,\right) & \left(\,80_{-20}^{+50}\,\right)~\mbox{[U18]} \\
ESO\,137-G034 \dotfill & >\!160 & \nodata & >\!150 & \nodata & \nodata & >\!230 & \nodata \\
{\bf NGC\,6300} \dotfill & 210_{-50}^{+100} & \nodata & >\!250 & >\!520 & \nodata & \left(\,80_{-20}^{+30}\,\right) & \nodata \\
2MASX\,J2018 \dotfill & 170_{-70}^{+u} & \nodata & \left(\,50_{-20}^{+110}\,\right) & \nodata & \nodata & >\!60 & \nodata \\
ESO\,103-G035 \dotfill & 100_{-10}^{+20} & \nodata & \left(\,50_{-30}^{+250}\,\right) & >\!30 & \nodata & \left(\,60_{-10}^{+20}\,\right) & 100_{-30}^{+90}~\mbox{[B18]} \\
Cyg\,A \dotfill & 130_{-30}^{+70} & \nodata & \nodata & \nodata & \nodata & \left(\,100_{-10}^{+20}\,\right) & >\!110~\mbox{[R15]} \\
MCG\,$+$04-48-002 \dotfill & >\!150 & \nodata & >\!180 & \nodata & \nodata & >\!60 & \nodata \\
{\bf NGC\,7172} \dotfill & >\!670 & >\!40 & \left(\,70\pm10\,\right) & \nodata & \nodata & \left(\,110\pm20\,\right) & \left(\,70\pm10\,\right)~\mbox{[R19]} \\
NGC\,7582 \dotfill & 200_{-80}^{+190} & >\!50 & \nodata & \nodata & \nodata & >\!130 & \nodata \\
2MASX\,J2330 \dotfill & >\!70 & \nodata & \left(\,60_{-40}^{+340}\,\right) & \nodata & \nodata & >\!90 & \nodata \\
PKS\,2331$-$240 \dotfill & 220_{-110}^{+u} & \nodata & \nodata & \nodata & \nodata & \nodata & >\!250~\mbox{[U18]} \\
PKS\,2356$-$61 \dotfill & 160_{-80}^{+u} & \nodata & \nodata & \nodata & \nodata & >\!50 & >\!60~\mbox{[U18]} \\
\enddata
\tablecomments{ Targets highlighted with bold font are discussed individually in the Appendix. Constraints in parentheses correspond to $\tau_e\!>\!3$, calculated using Equation~\ref{eq:tauline}. References: D07\,$=$\,\citet{dadina-2007}, dR12\,$=$\,\citet{deRosa+2012}, M13\,$=$\,\citet{molina+2013}, V13\,$=$\,\citet{vasudevan+2013}, P14\,$=$\,\citet{puccetti+2014-ngc4945}, B15\,$=$\,\citet{balokovic+2015-mcg5}, M15\,$=$\,\citet{matt+2015-ngc5506}, R15\,$=$\,\citet{reynolds+2015-cyga}, F16\,$=$\,\citet{fuerst+2016-cenA}, R17\,$=$\,\citet{ricci+2017-bass}, B18\,$=$\,\citet{buisson+2018-eso103}, U18\,$=$\,\citet{ursini+2018-obscuredRGs}, R19\,$=$\,\citet{rani+2019}, U19\,$=$\,\citet{ursini+2019-ngc2110+4388}. }
\end{deluxetable*}